\documentclass[aps, prb, reprint, superscriptaddress, notitlepage, letterpaper,  10pt, floatfix, showpacs, longbibliography, balancelastpage, nofootinbib]{revtex4-2}
\pdfoutput=1
\usepackage{amssymb, graphicx, makecell, color, xcolor, amsmath, bm, url, float, mathrsfs, braket, bbold, psfrag, dcolumn, enumitem, array, tikz, algpseudocode, setspace, adjustbox, tabularx, ragged2e, booktabs, silence, flushend, mdframed, stackengine, tcolorbox}
\setstackEOL{\\}
\usetikzlibrary{quantikz2}
\usepackage[normalem]{ulem}
\usepackage[mathlines]{lineno}
\usepackage[section]{placeins}
\usepackage[outline]{contour}
\usepackage[linesnumbered, ruled, vlined]{algorithm2e}
\usepackage[colorlinks = True, citecolor = blue, linkcolor = blue, anchorcolor=red, pdftoolbar = false, bookmarks = false]{hyperref}
\usepackage[utf8]{inputenc}
\usepackage{orcidlink}
\usepackage[symbol]{footmisc}

\begin{document}
\title{Resource-Efficient Hadamard Test Tailored Variational Framework for Nonlinear Dynamics on Quantum Computers}
\author{Eleftherios Mastorakis \orcidlink{0009-0005-3546-2568}}
\affiliation{School of Electrical and Computer Engineering, Technical University of Crete, Chania, Greece 73100}
\author{Muhammad Umer \orcidlink{0000-0002-1941-1833}}
\email{umer@u.nus.edu}
\affiliation{Centre for Quantum Technologies, National University of Singapore, 3 Science Drive 2, Singapore 117543}
\author{Milena Guevara-Bertsch}
\affiliation{Alpine Quantum Technologies GmbH, 6020 Innsbruck, Austria}
\author{Juris Ulmanis}
\affiliation{Alpine Quantum Technologies GmbH, 6020 Innsbruck, Austria}
\author{Felix Rohde}
\affiliation{Alpine Quantum Technologies GmbH, 6020 Innsbruck, Austria}
\author{Dimitris G. Angelakis \orcidlink{0000-0001-6763-6060}}
\email{dimitris.angelakis@gmail.com}
\affiliation{School of Electrical and Computer Engineering, Technical University of Crete, Chania, Greece 73100}
\affiliation{Centre for Quantum Technologies, National University of Singapore, 3 Science Drive 2, Singapore 117543}
\affiliation{AngelQ Quantum Computing, 531A Upper Cross Street, \#04-95 Hong Lim Complex, Singapore 051531}
\date{\today}

\begin{abstract}
Resource-efficient, low-depth implementations of quantum circuits remain a promising strategy for achieving reliable and scalable computation on quantum hardware, as they reduce gate resources and limit the accumulation of noisy operations. Here, we propose a low-depth implementation of a class of Hadamard test circuits, complemented by the development of a parameterized quantum ansatz specifically tailored for variational algorithms that exploit the underlying Hadamard test framework. Our findings demonstrate a significant reduction in single- and two-qubit gate counts, suggesting a reliable circuit architecture for noisy intermediate-scale quantum (NISQ) devices. Building on this foundation, we tested our low-depth scheme to investigate the expressive capacity of the proposed parameterized ansatz in simulating nonlinear Burgers\textquotesingle{} dynamics. The resulting variational quantum states faithfully capture the shockwave feature of the turbulent regime and maintain high overlaps with classical benchmarks, underscoring the practical effectiveness of our framework. Furthermore, we evaluate the effect of hardware noise by modeling the error properties of real quantum processors and by executing the variational algorithm on a trapped-ion-based \emph{IBEX Q1} device. The outcomes of our demonstrations highlight the resilience of our low-depth scheme in the turbulent regime, consistently preparing high-fidelity variational states that exhibit strong agreement with classical benchmarks. Our work contributes to the advancement of resource-efficient strategies for quantum computation, offering a robust framework for tackling a range of computationally intensive problems across numerous applications. 
\end{abstract}
\maketitle

\section{Introduction}
\label{Sec:Introduction}

\par Quantum computation has attracted notable interest in recent decades, as it offers the prospect of outstripping the capabilities of current supercomputers in tackling a range of classically intractable problems. In this regard, a broad spectrum of quantum algorithms has been developed, including Shor\textquotesingle{s} \cite{Shor1994} and Grover\textquotesingle{s} \cite{Grover1996} algorithms, the Harrow–Hassidim–Lloyd (HHL) solver \cite{Harrow2009}, quantum simulation protocols \cite{Abrams1997, Abrams1999}, variational quantum algorithms \cite{Cerezo2021, Bharti2022} and quantum machine learning methods \cite{Rebentrost2014, Lloyd2014, Havlivcek2019}, to name a few, to take advantage of quantum hardware to solve complex problems across diverse domains. Building on these algorithmic advances, attention has been devoted to applying these algorithms in quantum chemistry \cite{Peruzzo2014, OMalley2016, Kandala2017, Hempel2018, Ganzhorn2019, Quantum2020}, quantitative finance \cite{Cong2024, Huber2024, Sarma2024}, quantum dynamics \cite{Cirstoiu2020, Lin2021, Linteau2024}, computational fluid dynamics \cite{Lubasch2020, Jaksch2023}, combinatorial optimization \cite{Farhi2014, Pagano2020, Tan2021, Zhu2022}, and nonlinear physics \cite{Bengoechea2024, Siegl2025, Setty2025}, among other fields.

\par Various quantum algorithms, including the variational quantum algorithms \cite{Lubasch2020, Barison2022, Jaksch2023, Sarma2024, Yoshioka2024}, quantum phase estimation (QPE) \cite{Dobifmmode2007, Ding2024, Gunther2025}, Harrow-Hassidim-Lloyd (HHL) \cite{Harrow2009}, and quantum error correction frameworks \cite{Lidar2013, Devitt2013}, among others, employ the Hadamard test method for circuit construction to estimate state overlaps, evaluate expectation values, and perform syndrome measurements. Among its various advantages, the Hadamard test construction enables the extraction of computational outcomes solely by measuring the ancilla qubit, thereby minimizing the required readout resources. However, such quantum circuits typically suffer from large circuit depth and substantial gate overhead, posing significant challenges for their implementation on near-term quantum devices \cite{Umer2024}. Therefore, resource-efficient, low-depth implementations of Hadamard test circuits are crucial for the successful demonstration of these algorithms on (post-) NISQ-era hardware platforms.

\par Recently, efforts have been directed towards developing low-depth implementations of quantum circuits \cite{Chee2023}, including ancilla-free methods \cite{Clinton2024, Yang2024, Wang2025, Kastner2024} and resource-optimized variants of Hadamard-test-based schemes \cite{Wu2021, Tazi2024, Faehrmann2025}, in order to enhance their compatibility with (post-) NISQ devices. Ancilla-free approaches \cite{Clinton2024, Yang2024, Wang2025, Kastner2024} eliminate controlled unitaries and ancilla qubits altogether, thereby shifting the computational overhead to classical post-processing. By contrast, variants of Hadamard-test-based methods \cite{Wu2021, Tazi2024} intentionally increase the number of ancilla qubits to enable parallelization and reduce circuit depth, while keeping the measurement overhead minimal despite the increased circuit width. These variations of Hadamard-test methods are typically designed to achieve a more favourable scaling of the overall depth–width product compared to conventional Hadamard-test implementations \cite{Tazi2024}. However, it remains highly desirable to reduce the circuit depth without incurring an increase in circuit width, thereby enabling more resource-efficient implementations of Hadamard test circuits that are better aligned with the constraints of near-term quantum hardware. Here, we directly address this challenge by proposing an efficient realization of a broad class of Hadamard test quantum circuits, significantly reducing gate resource requirements. 

\par In this article, our objective is two-fold. First, we present a low-depth implementation of a class of Hadamard test circuits while preserving the computational integrity of the original construction. In particular, as we shall demonstrate, our approach replaces certain conditional quantum gates with $N$ controls by equivalent gates with only $N-1$ controls, thereby reducing the required computational resources and the overall depth of the quantum circuits. For example, in the case of a Toffoli gate, our construction effectively reduces a three-qubit controlled operation to a single CNOT gate, thereby saving on the order of five CNOT gates (for a standard decomposition) together with multiple single-qubit rotations, and thus yielding a substantially shallower implementation. In addition to this, we propose an ansatz structure specifically tailored to the Hadamard test framework. To evaluate the effectiveness of our proposed low-depth constructions, we investigate the nonlinear dynamics of fluid fields governed by the Burgers\textquotesingle{} equation and demonstrate a substantial reduction in two-qubit gate count across various basis gate sets and qubit connectivity configurations. Second, we execute the variational algorithm on the trapped-ion-based \emph{IBEX Q1} quantum device and examine the impact of hardware noise on the resource-efficient evaluation of the cost function and the underlying nonlinear Burgers\textquotesingle{} dynamics. It is worth emphasizing that the encoding and representation of a broad class of nonlinear physics and computational fluid dynamics problems on quantum computers have already been the focus of numerous prior studies \cite{Lubasch2020, Jaksch2023, Umer2024, Bengoechea2024, Umer2024opt}. Our analysis demonstrates that the time evolution of fluid fields in the turbulent regime, characterized by strongly nonlinear dynamics and steep gradient or shock-like features, is resolved with high fidelity on a NISQ-era quantum device for small-scale problem instances, such as those discretized on a spatial grid of $2^3$ points and evolved for a couple of time steps. Notably, to the best of our knowledge, this constitutes the first successful observation of turbulent fluid dynamics captured using variational methods on current trapped-ion hardware. Our findings highlight the efficacy of low-depth Hadamard test construction in harnessing NISQ devices for meaningful quantum simulations.

\par The remainder of this article is structured as follows. Section~\ref{Sec:Low_Depth_Hadamard} introduces a generic low-depth implementation of Hadamard test circuits. In particular, we present a general framework that is independent of the specific problem instance and relies solely on the structural features of the Hadamard test construction. Subsequently, in section~\ref{Sec:Hadamard_Ansatz}, we propose a tailored ansatz structure designed specifically for circuits based on the Hadamard test framework. In section~\ref{Sec:Burgers_Dynamics}, we implement our proposed low-depth construction in nonlinear dynamics problem of Burgers\textquotesingle{} equation. Here, we also analyze the expressivity of the proposed ansatz in noiseless settings. In addition to this, section~\ref{Sec:Gate_Reduction} discusses circuit depth reduction in terms of two-qubit gate count. In section ~\ref{Sec:Noisy_Simulations}, we evaluate the practical applicability of our low-depth scheme to simulate Burgers\textquotesingle{} dynamics under hardware-noise models derived from multiple quantum processors. Guided by the outcomes of the noisy simulations, we then implement the algorithm on the AQT\textquotesingle{s} trapped-ion-based \emph{IBEX-Q1} device and report the empirical results in section~\ref{Sec:Device_Simulations}. Finally, we summarize in section~\ref{Sec:Summary} and present an outlook for future research directions. 

\section{Low-Depth Implementation of the Hadamard Test Quantum Circuits}
\label{Sec:Low_Depth_Hadamard}

\par Simulating and benchmarking complex quantum algorithms in the NISQ era is a challenging task, often constrained by the significant noise levels inherent in present devices \cite{Chee2024, Umer2024}. Device noise typically affects various stages of quantum computation \cite{Nielsen2010}, including the initialization process (reset noise), gate operations (gate noise), and the measurement process (readout noise), to name a few. One approach to achieving reliable, scalable quantum computation is to employ low-depth quantum circuits, thereby reducing the number of noise-prone gate operations \cite{Chee2023}. Depth minimization strategy remains advantageous even within fault-tolerant architectures \cite{Niemann2019, Chatterjee2025}.

\par
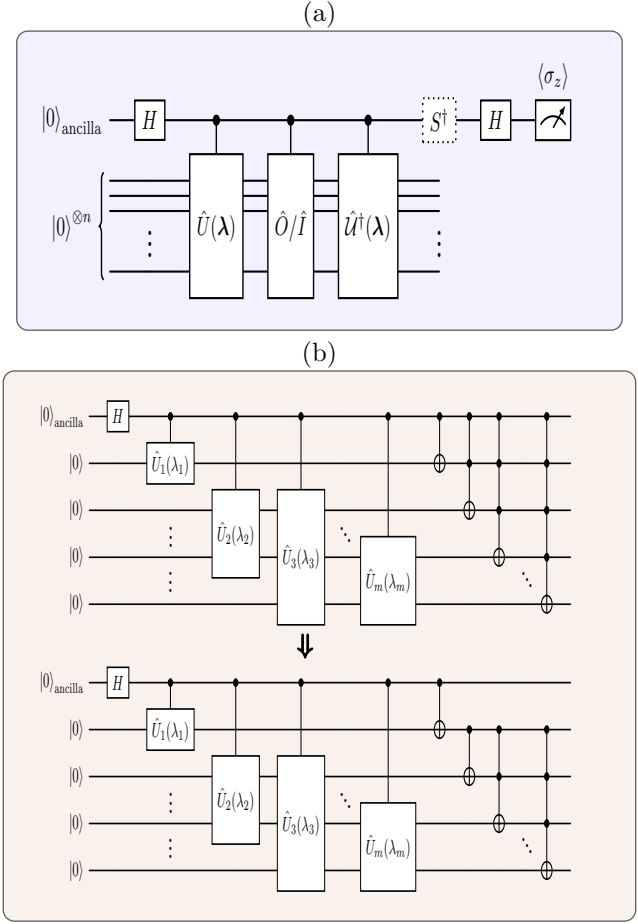
\begin{figure}[htb]
    \centering
    \sbox0{
        \begin{adjustbox}{width = 0.40\textwidth, height = 0.11\textwidth}
        \begin{quantikz}[row sep={0.8cm,between origins}, wire types={q, n, q, q, q}]
        \lstick{$\ket{0}_{\rm ancilla}$} & \gate{H}\gategroup[1, steps = 1, style = {draw=none, fill=none, inner xsep = 0pt}, background, label style={label position = above, anchor = north, yshift = -1.9cm, xshift = -0.0cm}]{\boldsymbol{\vdots}} & \ctrl{1} & \ctrl{1} & \ctrl{1} & \gate[style={draw, dotted}]{S^{\dagger}}\gategroup[1, steps = 1, style = {draw=none, fill=none, inner xsep = 0pt}, background, label style={label position = above, anchor = north, yshift = -1.9cm, xshift = 0.0cm}]{\boldsymbol{\vdots}} & \gate{H}& \meter{\langle\sigma_{z}\rangle} \\ 
        \lstick[4, braces = {black, thick}]{$\ket{0}^{\otimes{n}}$} & \setwiretype{q} & \gate[4]{\hat{U}({\boldsymbol\lambda})} & \gate[4]{\hat{O}/\hat{I}} & \gate[4]{\hat{\mathcal{U}}^{\dagger}({\boldsymbol\lambda})} & \\ [-0.6cm]
        & & & & & \\ [-0.6cm]
        & & & & & \\ 
        & & & & &
        \end{quantikz}
        \end{adjustbox}
        }
    \sbox1{
        \begin{adjustbox}{width = 0.40\textwidth, height = 0.09\textwidth}
        \begin{quantikz}[row sep={0.8cm,between origins}, wire types={q, q, q, q, q}]
        \lstick{$\ket{0}_{\rm ancilla}$} & \gate{H} & \ctrl{1} & \ctrl{2} & \ctrl{2} & & \ctrl{3} & \ctrl{1} & \ctrl{2} & \ctrl{3} & & \ctrl{4} & \\
        \lstick{$\ket{0}$} & & \gate[1]{\hat{U}_{1}(\lambda_{1})}\gategroup[1, steps = 1, style={draw=none, fill=none, inner xsep = 0pt}, background, label style={label position = above, anchor = north, yshift = -2.35cm, xshift = -0.0cm}]{\boldsymbol{\vdots}} & & & & & \targ{} & \control{} & \control{} & & \control{} &\\ 
        \lstick{$\ket{0}$} & & & \gate[2]{\hat{U}_{2}(\lambda_{2})}\gategroup[1, steps = 1, style={draw=none, fill=none, inner xsep = 0pt}, background, label style={label position = above, anchor = north, yshift = -0.78cm, xshift = -1.8cm}]{\boldsymbol{\vdots}} & \gate[3]{\hat{U}_{3}(\lambda_{3})}\gategroup[1, steps = 1, style={draw=none, fill=none, inner xsep = 0pt}, background, label style={rotate = 45, label position = above, anchor = north, yshift = -1.25cm, xshift = 0.0cm}]{\boldsymbol{\vdots}} & & & & \targ{} & \control{} & & \control{} & \\
        \lstick{$\ket{0}$} & & & & & & \gate[2]{\hat{U}_{m}(\lambda_{m})} & & & \targ{}\gategroup[1, steps = 1, style={draw=none, fill=none, inner xsep = 0pt}, background, label style={rotate = 45, label position = above, anchor = north, yshift = -0.9cm, xshift = -0.3cm}]{\boldsymbol{\vdots}} & & \control{} & \\
        \lstick{$\ket{0}$} & & & & & & & & & & & \targ{} &
        \end{quantikz}
        \end{adjustbox}
        }
    \sbox2{
        \begin{adjustbox}{width = 0.40\textwidth, height = 0.09\textwidth}
        \begin{quantikz}[row sep={0.8cm,between origins}, wire types={q, q, q, q, q}]
        \lstick{$\ket{0}_{\rm ancilla}$} & \gate{H} & \ctrl{1} & \ctrl{2} & \ctrl{2} & & \ctrl{3} & \ctrl{1} & & & & &  \\
        \lstick{$\ket{0}$} & & \gate[1]{\hat{U}_{1}(\lambda_{1})}\gategroup[1, steps = 1, style={draw=none, fill=none, inner xsep = 0pt}, background, label style={label position = above, anchor = north, yshift = -2.35cm, xshift = -0.0cm}]{\boldsymbol{\vdots}} & & & & & \targ{} & \ctrl{1} & \ctrl{2} & & \ctrl{3} &\\
        \lstick{$\ket{0}$} & & & \gate[2]{\hat{U}_{2}(\lambda_{2})}\gategroup[1, steps = 1, style={draw=none, fill=none, inner xsep = 0pt}, background, label style={label position = above, anchor = north, yshift = -0.78cm, xshift = -1.8cm}]{\boldsymbol{\vdots}} & \gate[3]{\hat{U}_{3}(\lambda_{3})}\gategroup[1, steps = 1, style={draw=none, fill=none, inner xsep = 0pt}, background, label style={rotate = 45, label position = above, anchor = north, yshift = -1.25cm, xshift = 0.0cm}]{\boldsymbol{\vdots}} & & & & \targ{} & \control{} & & \control{} & \\
        \lstick{$\ket{0}$} & & & & & & \gate[2]{\hat{U}_{m}(\lambda_{m})} & & & \targ{}\gategroup[1, steps = 1, style={draw=none, fill=none, inner xsep = 0pt}, background, label style={rotate = 45, label position = above, anchor = north, yshift = -0.9cm, xshift = -0.3cm}]{\boldsymbol{\vdots}} & & \control{} & \\
        \lstick{$\ket{0}$} & & & & & & & & & & & \targ{} &
        \end{quantikz}
        \end{adjustbox}
        }
    \begin{tabular}{c}
    (a) \\ 
    \begin{tcolorbox}[width = 0.45\textwidth, colback = blue!5!white, colframe = gray!75!black, boxrule = 0.5pt, arc = 4pt, halign = center]
    \hspace*{-0.38cm}\usebox0 
    \end{tcolorbox} \\
    (b) \\
    \begin{tcolorbox}[width = 0.47\textwidth, colback = brown!10!white, colframe = gray!75!black, boxrule = 0.5pt, arc = 4pt, halign = center]
    \hspace*{-0.38cm}\usebox1 \\ \hspace*{-0.38cm}\contour{black}{$\Downarrow$} \\ \hspace*{-0.38cm}\usebox2
    \end{tcolorbox}
    \end{tabular}
    \caption{Low-depth implementation of the Hadamard test circuits. Panel (a) shows the general construction of the Hadamard test circuits used to evaluate matrix elements or state overlaps. Here, the absence (presence) of the $S^{\dagger}$ gate allows the estimation of the real (imaginary) component of the complex amplitude. In panel (b), we illustrate our low-depth construction, achieved by systematically omitting redundant operations from the original quantum circuit. \vspace{-0.2cm}}
    \label{Fig:Hadamard_Test}
\end{figure}

 Here, we propose a resource-efficient, low-depth implementation scheme for a class of quantum circuits based on the Hadamard test. A standard Hadamard-test circuit (i) implements each unitary operation controlled by an ancilla qubit, (ii) places the controlled operations between two Hadamard gates acting on the ancilla qubit, and (iii) extracts the circuit outcome from a projective measurement of the ancilla qubit, as illustrated in Fig.~\ref{Fig:Hadamard_Test}a. It is important to note that controlling each unitary operation from the ancilla qubit yields multi-qubit operations, including Toffoli and multi-controlled Toffoli gates. When these multi-qubit operations are decomposed into the single- and two-qubit unitary operations available in the hardware\textquotesingle{s} native gate set \cite{Nielsen2010}, the resulting circuits become markedly deeper and incur a higher count of noise-prone two-qubit gate operations.

\par In conjunction with the aforementioned Hadamard circuit architecture, our approach imposes a single constraint: each qubit must be initialized in the $\ket{0}$ state. This constraint is straightforward to satisfy and remains the standard initialization choice on most contemporary hardware platforms \cite{IBMQ, AQT_IBEX, Pogorelov2021}. As we demonstrate below, our proposed low-depth construction of the Hadamard test quantum circuits can be summarized as follows:
\begin{center}
  \begin{minipage}{0.46\textwidth}
    \emph{\vspace{-0.00cm} Given the Hadamard test framework, together with the initialization of every qubit in the $\ket{0}$ state, any quantum gate operation with at least one control on a non-ancilla qubit does not require a control from the ancilla qubit. \vspace{0.1cm}}
  \end{minipage}
\end{center}
To substantiate the above-mentioned claim, we consider the generic Hadamard test circuit shown in Fig.~\ref{Fig:Hadamard_Test}a, which is routinely employed to evaluate either matrix elements $\braket{0|\hat{\mathcal{U}}({\boldsymbol\lambda})\hat{O}\hat{U}({\boldsymbol\lambda})|0}$ of some arbitrary operator $\hat{O}$ or state overlaps $\braket{0|\hat{\mathcal{U}}({\boldsymbol\lambda})\hat{U}({\boldsymbol\lambda})|0}$. Following the execution of the controlled-$\hat{U}({\boldsymbol\lambda})$ operation, the qubit system resides in the state $\ket{\Psi_{H}}$ expressed as
\begin{eqnarray}\begin{aligned}
\label{EQ:Hadamard}
\ket{\Psi_{H}} = {\rm C}^{1}\hat{U}({\boldsymbol\lambda})\Bigl[\frac{1}{\sqrt{2}}\bigl[\ket{0}^{\otimes{n}}\ket{0}_{a} + \ket{0}^{\otimes{n}}\ket{1}_{a}\bigl]\Bigl]\;. 
\end{aligned}\end{eqnarray}
Here, $\ket{\cdots}^{\otimes n}$ ($\ket{\cdots}_{a}$) represent the state of the $n$-qubit register (ancilla qubit) and ${\rm C}^{1}\hat{U}({\boldsymbol\lambda})$ is a controlled unitary conditioned on a control qubit. Without loss of generality, we consider a representative form of the unitary operator $\hat{U}({\boldsymbol\lambda})$, comprising parameterized single- and multi-qubit rotation operators together with CNOT and various multi-controlled Toffoli gates, as illustrated in Fig.~\ref{Fig:Hadamard_Test}b $\bigl($up to a control from the ancilla qubit$\bigr)$. The controlled parameterized operators $\hat{U}_{i}(\lambda_{i})$ act non-trivially only on the subspace in which the ancilla qubit is in the state $\ket{1}_{a}$, such that Eq.~(\ref{EQ:Hadamard}) takes the form 
\begin{flushleft}
\begin{eqnarray}\begin{aligned}
\label{EQ:Hadamard1}
\ket{\Psi_{H}} =& \hat{\mathcal{O}}\prod_{j=1}^{m'}{\rm C}^{p}{\rm NOT}(a, \{q_{k_{1}}, \dots, q_{k_{p-1}}\}, q_{j})\Bigl[\frac{1}{\sqrt{2}}\bigl[\ket{0}^{\otimes{n}}\ket{0}_{a} \\ &~~~+ \hat{\mathcal{O}}\prod_{i = 1}^{m}\hat{U}_{i}(\lambda_{i})\ket{0}^{\otimes{n}}\ket{1}_{a}\bigl]\Bigl]\;, 
\end{aligned}\end{eqnarray}
\end{flushleft}
where $\hat{\mathcal{O}}$ denotes an ordering operator, ${a}$ labels the ancilla qubit, and $\{q_{k_{1}}, \dots, q_{k_{p-1}}\}$ designates the $p-1$ control qubits within the $n$-qubit register (hereafter referred to as the intrinsic control qubits), with indices satisfying $1 \leq k_{1} \leq \dots \leq k_{p-1} \leq n$. $\hat{U}_{i}(\lambda_{i})$ is a parameterized or fixed unconditional unitary operator, ${\rm C}^{p}{\rm NOT}$ denotes a conditional gate with $p$ control qubits $(a, \{q_{k_{1}}, \dots, q_{k_{p-1}}\})$ and target qubit $q_{j}$ on which the NOT ($X$) operation is applied. It is worth noting that for all conditional operations of the form ${\rm C^{p}NOT}(a, \{q_{k_{1}}, \dots, q_{k_{p-1}}\}, q_{j})$, which are intrinsically controlled by the qubits $\{q_{k_{1}}, \dots, q_{k_{p-1}}\}$, the ancilla control is redundant and does not alter the gate\textquotesingle{s} operation. This observation can be clarified as follows. (i) For the $\ket{0}_{a}$ subspace, both the ancilla and the intrinsic control qubits remain in the trivial state, so the conditional operator leaves this subspace unchanged. (ii) Within the $\ket{1}_{a}$ subspace, the ancilla qubit is always nontrivial, and the action of the conditional operation ${\rm C}^{p}{\rm NOT}(a, \{q_{k_{1}}, \dots, q_{k_{p-1}}\}, q_{j})$ remains determined exclusively by the intrinsic control set $\{q_{k_{1}}, \dots, q_{k_{p-1}}\}$. Consequently, the ancilla control is redundant, and the conditional operation effectively reduces to ${\rm C}^{p-1}{\rm NOT}(\{q_{k_{1}}, \dots, q_{k_{p-1}}\}, q_{j})$, i.e., a $(p-1)$-controlled NOT gate. As a result, Eq.~(\ref{EQ:Hadamard1}) takes the form
\begin{flushleft}
\begin{eqnarray}\begin{aligned}
\label{EQ:Hadamard2}
\ket{\Psi_{H}} =& \frac{1}{\sqrt{2}}\Bigl[\ket{0}^{\otimes{n}}\ket{0}_{a} + \prod_{j=2}^{m'}{\rm C}^{p-1}{\rm NOT}(\{q_{k_{1}}, \dots, q_{k_{p-1}}\}, q_{j})\\ &~~~~~~~{\rm C}^{1}{\rm NOT}(a, q_{1})\prod_{i = 1}^{m}\hat{U}_{i}(\lambda_{i})\ket{0}^{\otimes{n}}\ket{1}_{a}\Bigl]\;. 
\end{aligned}\end{eqnarray}
\end{flushleft}
It is worth highlighting that the conditional gate ${\rm C}^{1}{\rm NOT}(a, q_{1})$, which lacks intrinsic controls, necessarily retains its ancilla control, as illustrated in Fig.~\ref{Fig:Hadamard_Test}b and Eq.~(\ref{EQ:Hadamard2}). The implementation of these $(p-1)$-controlled operations requires substantially fewer native basis gates, compared to $p$-controlled gates, on any given quantum processor. For example, eliminating redundant ancillary control turns a three-qubit Toffoli gate into a single CNOT, thereby lowering the gate resources by five two-qubit gates and at least nine single-qubit operations.

\par This simple mathematical observation yields significantly shallower circuits, thereby enhancing the practical feasibility of implementing Hadamard test circuits. In architectures without all-to-all qubit connectivity, eliminating the ancilla control from multi-controlled CNOT gates markedly reduces the swap-gate overhead needed to realize some conditional operations between non-adjacent qubits. In what follows, we propose an ansatz structure to mitigate the gate resource overhead incurred by ancilla-controlled parameterized unitary operations within the variational settings.

\begin{figure*}[hbt]
    \centering
    \sbox3{
        \begin{adjustbox}{width = 0.80\textwidth, height = 0.115\textwidth}
        \begin{quantikz}[row sep={0.8cm,between origins}, wire types={q,q,q,q,q}]
        \lstick{$\ket{0}_{a}$} & \gate{H} & \ctrl{1} & & \push{\hspace{2cm}} & \gate{H} & \ctrl{1} & & & & & & & & \\ [-0.1cm]
        \lstick{$\ket{0}_{q_{1}}$} & & \gate[1]{\hat{X}~{\rm or}~\hat{\tilde{U}}(\lambda_{0})} & \gate[4]{~~\hat{\mathbf{U}}(\boldsymbol{\lambda})~~}\gategroup[1, steps = 1, style = {draw=none, fill=none}, background, label style={label position = above, anchor = north, yshift = -1.35cm, xshift = 2.1cm}]{{\Large $\equiv$}} & \push{\hspace{2cm}} & & \gate[1]{\hat{X}~{\rm or}~\hat{\tilde{U}}(\lambda_{0})} & \ctrl{1}\gategroup[4, steps = 4, style = {dashed, rounded corners, fill=brown!10, inner xsep = 2pt}, background, label style={label position = above, anchor = north, yshift = -0.2cm, xshift = 0.0cm}]{layer} & & & \gate[1]{\hat{\tilde{U}}(\lambda_{l})} & & & & \\ [-0.1cm]
        \lstick{$\ket{0}_{q_{2}}$} &  & & & \push{\hspace{2cm}} & & & \gate[1]{\hat{\tilde{U}}(\lambda_{i})} & \ctrl{1} & & & & & & \\ [-0.1cm]
        \lstick{$\ket{0}_{q_{3}}$} &  & & & \push{\hspace{2cm}} & & & & \gate[1]{\hat{\tilde{U}}(\lambda_{j})} & \ctrl{1}\gategroup[1, steps = 1, style = {draw=none, fill=none, inner xsep = 2pt}, background, label style={label position = above, anchor = north, yshift = -0.17cm, xshift = 4.0cm}]{$\boldsymbol{\cdots}$} & & & & & \\ [-0.1cm]
        \lstick{$\ket{0}_{q_{4}}$} &  & & & \push{\hspace{2cm}} & & & & & \gate[1]{\hat{\tilde{U}}(\lambda_{k})} & \ctrl{-3}\gategroup[1, steps = 1, style = {draw=none, fill=none, inner xsep = 2pt}, background, label style={label position = above, anchor = north, yshift = 0.35cm, xshift = 2.20cm}]{$d-1$} & & & &
        \end{quantikz}
        \end{adjustbox}
        }
    \sbox4{
        \begin{adjustbox}{width = 0.40\textwidth, height = 0.045\textwidth}
        \begin{quantikz}[row sep={0.8cm,between origins}, wire types={q,q}]
        \lstick{~} & \ctrl{1} & \push{\hspace{2cm}} & \ctrl{1} & & \ctrl{1} & &\\ [-0.1cm]
        \lstick{~} & \gate[1]{\hat{\tilde{U}}(\lambda)}\gategroup[1, steps = 1, style = {draw=none, fill=none}, background, label style={label position = above, anchor = north, yshift = -0.35cm, xshift = 2.0cm}]{$\equiv$} & \push{\hspace{2cm}} & \targ{} & \gate[1]{\hat{\tilde{U}}(-\lambda/2)} & \targ{} & \gate[1]{\hat{\tilde{U}}(\lambda/2)} &
        \end{quantikz}
        \end{adjustbox}
        }
    \sbox5{
        \begin{adjustbox}{width = 0.40\textwidth, height = 0.045\textwidth}
        \begin{quantikz}[row sep={0.8cm, between origins}, wire types={q,q}]
        \lstick{~} & \ctrl{1} & \push{\hspace{2cm}} & & & \ctrl{1} & &\\ [-0.1cm]
        \lstick{~} & \gate[1]{\hat{\overline{U}}(\lambda)}\gategroup[1, steps = 1, style = {draw=none, fill=none}, background, label style={label position = above, anchor = north, yshift = -0.35cm, xshift = 2.0cm}]{$\equiv$} & \push{\hspace{2cm}} & & \gate[1]{\hat{\overline{U}}_{s}(-\lambda/2)} & \targ{} & \gate[1]{\hat{\overline{U}}_{s}(\lambda/2)} &
        \end{quantikz}
        \end{adjustbox}
        }
    \begin{tabular}{cc}
    \multicolumn{2}{c}{(a)}\\ 
    \multicolumn{2}{c}{\usebox3}\vspace{0.3cm}\\
    (b) & ~~(c) \\ 
    \begin{tcolorbox}[width = 0.48\textwidth, colback = blue!10!white, colframe = gray!75!black, boxrule = 0.5pt, arc = 4pt, halign = center]
    \hspace*{-0.3cm}\usebox4
    \end{tcolorbox}
    & ~~\begin{tcolorbox}[width = 0.48\textwidth, colback = gray!10!white, colframe = gray!75!black, boxrule = 0.5pt, arc = 4pt, halign = center]
    \hspace*{-0.3cm}\usebox5
    \end{tcolorbox}
    \end{tabular}
    \caption{ Ansatz structure tailored to the Hadamard test construction. Panel (a) shows the ansatz structure with only controlled-$\hat{\tilde{U}}(\lambda_{i})$ operations between the qubits. The ansatz consists of $d$ number of layers where the first layer is highlighted with the beige color. Panel (b) shows one possible controlled-$\hat{\tilde{U}}(\lambda_{i})$ gate, where $\hat{\tilde{U}} \in \{\hat{R}_{x}, \hat{R}_{y}, \hat{R}_{z}\}$. Here, each controlled-$\hat{\tilde{U}}(\lambda_{i})$ gate decomposes into two controlled-NOT gates and two single-qubit rotations. Panel (c) shows an alternate selection for the controlled-$\hat{\overline{U}}(\lambda_{i})$ operation that only consist of one CNOT gate and two single-qubit rotations $\hat{\overline{U}}_{s}(\pm\lambda/2)$. It is worth highlighting that the panel (b) and (c) show different selections of controlled-unitary operations, which are not equivalent to each other. \vspace{-0.2cm} }
    \label{Fig:Proposed_Ansatz}
\end{figure*}
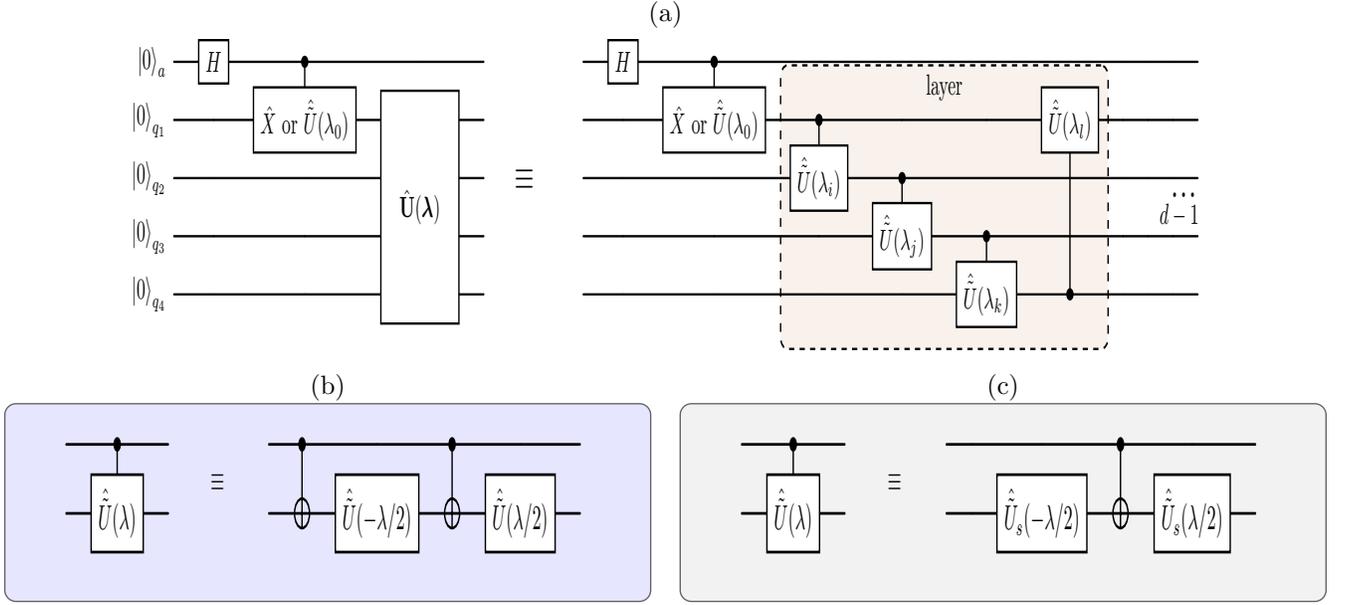

\subsection{Ansatz Structure tailored to the Hadamard test construction}
\label{Sec:Hadamard_Ansatz}

\par In this section, we build upon our understanding of Hadamard test constructions and propose a novel ansatz structure designed to significantly reduce both the circuit depth and the number of noisy two-qubit gates in variational quantum algorithms. The proposed architecture is motivated by the key observation that conditional gates with at least one control qubit within the main quantum register can be implemented without requiring an ancilla control. To this end, and in order to eliminate the need for control operations from the ancilla qubit, we restrict the ansatz structure to include only conditional parameterized gates, as illustrated in Fig.~\ref{Fig:Proposed_Ansatz}a. As an initial step, we apply either a controlled-NOT or a controlled-$\hat{\tilde{U}}$ operation from the ancilla qubit to one of the qubits in the main quantum register. This ancilla-controlled operation establishes correlations between the ancilla and the quantum register, thereby enabling selective application of subsequent conditional parameterized gates based solely on the state of the register qubits. One such gate sequence is depicted in Fig.~\ref{Fig:Proposed_Ansatz}a. In this configuration, the controlled-$\hat{\tilde{U}}$ gates serve two primary functions: (i) facilitate entanglement among the qubits in the register, and (ii) enable parameterization of the variational quantum state. It is worth emphasizing that the control qubits for the controlled-$\hat{\tilde{U}}$ gate are confined to those qubits that have previously served as targets in earlier controlled operations. By appropriately selecting the form of the unitary operator $\hat{\tilde{U}}$ and tuning the number of layers according to the requirements of the problem, the ansatz is expected to effectively represent the target solution space. This design inherently eliminates control operations from the ancilla qubit for the parameterized quantum circuit and thereby achieves a substantial reduction in the total two-qubit gate count, relative to the ancilla-controlled ansatz illustrated in Fig.~\ref{Fig:Hadamard_Test}a-\ref{Fig:Hadamard_Test}b. 

\par A few comments are in order. First, the initial ancilla-controlled NOT or $\hat{\tilde{U}}$ operation induces a nontrivial transformation within the $\ket{1}_{a}$ subspace, thereby establishing the necessary entanglement between the ancilla and the quantum register. It turns out that this is the only interaction required between the ancilla qubit and the remainder of the qubits for our proposed ansatz structure. All subsequent operations remain confined to the register qubits, leveraging conditional parameterized gates to explore the $n$-qubit Hilbert space in pursuit of the variational solution. This ansatz significantly reduces the cumulative two-qubit gate overhead by exploiting the underlying logical structure of the Hadamard test circuits.

\par Second, although we have restricted ourselves to controlled-$\hat{\tilde{U}}$ operations that are readily implementable on current quantum hardware, namely $\hat{\tilde{U}} \in {\hat{R}_x, \hat{R}_y, \hat{R}_z}$ and the unitary configuration illustrated in Fig.~\ref{Fig:Proposed_Ansatz}c, this selection is by no means exhaustive. For instance, problem-inspired ansatz structures can be developed and tailored for specific applications by selecting unitary operations and entanglement patterns \cite{Note1} that encode domain-specific knowledge. It is worth emphasizing that both the choice of $\hat{\tilde{U}}$ and the associated entanglement pattern critically influence the expressivity of the ansatz. Exploring problem-informed extensions of this design framework remains a promising direction for future research.

\par Finally, a notable advantage of the proposed ansatz structure lies in its compatibility with quantum hardware platforms that exhibit nearest-neighbor (limited) qubit connectivity. In such hardware architectures, the controlled-$\hat{U}(\lambda_{i})$ operations of Fig.~\ref{Fig:Hadamard_Test}b often necessitate multiple swap gates to bring distant qubits into proximity, thereby increasing circuit depth and error susceptibility. In contrast, the controlled-$\hat{\tilde{U}}(\lambda_{i})$ operations of our proposed ansatz (Fig.~\ref{Fig:Proposed_Ansatz}a) are deliberately applied between adjacent qubits, enabling implementation with minimal swap overhead. This leads to significantly shallower circuits that are better suited to noisy intermediate-scale quantum (NISQ) devices. Consequently, the ansatz offers a hardware-efficient alternative aligned with the architectural constraints of some quantum processors.

\section{Nonlinear Burgers\textquotesingle{} Dynamics}
\label{Sec:Burgers_Dynamics}

\par The low-depth implementation of the Hadamard test circuits discussed in Sec.~\ref{Sec:Low_Depth_Hadamard} is generic, and can be applied for applications in quantum chemistry \cite{Peruzzo2014, Kandala2017, Wecker2015}, quantum dynamics \cite{Cirstoiu2020, Lin2021, Linteau2024}, quantum finance \cite{Sarma2024}, and combinatorial optimization problems \cite{Farhi2014, Tan2021}. In this section, however, we concentrate on probing the nonlinear dynamics of fluid velocity fields through our proposed implementation of Hadamard test circuits. In this regard, we consider one-dimensional Burgers\textquotesingle{} equation
\begin{eqnarray}\begin{aligned}
\label{Eq:Burgers}
\partial_{t}{u(x,~t)} = \nu\partial^2_{x} {u(x,~t)} - {u(x,~t)}\partial_{x}{u(x,~t)} \;,
\end{aligned}\end{eqnarray}
where $u(x,~t)$ is the fluid velocity field, $\nu$ denotes the coefficient of kinematic viscosity that captures the turbulent ($\nu < 1$) and the laminar ($\nu \geq 1$) regimes, $x$ ($t$) represents space (time), and $\partial_{x}$ ($\partial_{t}$) is the space (time) differential operation. 

\par Although the formulation of Burgers\textquotesingle{} dynamics on quantum processors has been discussed in previous works \cite{Lubasch2020, Jaksch2023, Umer2024opt}, we briefly revisit the procedure here for the sake of completeness. Consistent with standard numerical approaches, we discretize the spatial domain $x \in [a,~b]$ and temporal interval $t \in [0,~T]$ into uniformly spaced grid points, yielding discrete coordinates $x_{n_{x}} = a + \delta_{x}{n_{x}}$ and $t_{n_{t}} = \tau{n_{t}}$, with spacings $\delta_{x} = (b - a)/N_{x}$ and $\tau = T/N_{t}$. Considering periodic boundary conditions in the spatial direction, such that $u(x = a) = u(x = b)$, the velocity field values at a given time $t$ are collectively represented as a vector $\ket{{\boldsymbol{u}}_{t}}$. Since Burgers\textquotesingle{} equation does not preserve the vector norm $\braket{{\boldsymbol{u}}_{t}|{\boldsymbol{u}}_{t}}$, we introduce a hyperparameter $\Lambda_{t}$, relating it to a normalized quantum state via $\ket{{\boldsymbol{u}}_{t}} = \Lambda_{t}\ket{\Psi_{t}} = \Lambda_{t}\sum_{n_{x}}\psi_{n_{x}, t}\ket{{\rm binary}(n_{x})}$, with $\braket{\Psi_{t}|\Psi_{t}} = 1$. Furthermore, the spatial and temporal derivatives in Eq.~(\ref{Eq:Burgers}) are approximated using the central finite difference and the Euler methods, respectively, resulting in the following form:
\begin{eqnarray}\begin{aligned}
\label{EQ:Burgers1}
&\Lambda_{t+\tau}\ket{\Psi_{t+\tau}({\boldsymbol\lambda})} \\ & ~~~~~~~~ = \Bigl[\Lambda_{t} + l_{1}(\hat{A} + \hat{A}^{\dagger} - 2\hat{I}) - l_{2}\hat{D}_{t}(\hat{A} - \hat{A}^{\dagger})\Bigl]\ket{\Psi_{t}} \;, 
\end{aligned}\end{eqnarray}
where $\partial_{x}^{2} = \frac{\hat{A} + \hat{A}^{\dagger} - 2\hat{I}}{2\delta_{x}^{2}}$, $\partial_{x} = \frac{\hat{A} - \hat{A}^{\dagger}}{2\delta_{x}}$, $l_{1} = {\Lambda_{t}\tau\nu}/{2\delta_{x}^{2}}$, $l_{2} = {\vert\Lambda_{t}\vert^{2}\tau}/{2\delta_{x}}$, and $\hat{D}_{t} = {\rm diag}(\psi_{n_{x}, t})$. Here, $\hat{A}$ represents the adder operator which shifts the basis states by one unit \cite{Lubasch2020, Jaksch2023}, thereby implementing the spatial differentiation in the finite difference approximation. The cost function, defined as the squared residual of the variational state $\Lambda_{t+\tau}\ket{\Psi_{t+\tau}({\boldsymbol\lambda})}$ and the time-evolved state $\bigl[\Lambda_{t} + l_{1}(\hat{A} + \hat{A}^{\dagger} - 2\hat{I}) - l_{2}\hat{D}_{t}(\hat{A} - \hat{A}^{\dagger})\bigl]\ket{\Psi_{t}}$ is then given as
\begin{eqnarray}\begin{aligned}
\label{EQ:Burgers2}
C_{I}({\boldsymbol\lambda}) &= - \Bigl[\bigl(\Lambda^{*}_{t} - 2l_{1}^{*}\bigl)~{\rm Re}\{\bra{0} \hat{U}_{t}^{\dagger}\hat{U}_{t + \tau}({\boldsymbol\lambda})\ket{0}\} \\ 
& ~~~~~~~+ l_{1}^{*}{\rm Re}\{\bra{0} \hat{U}_{t}^{\dagger}\bigl(\hat{A} + \hat{A}^{\dagger}\bigl)\hat{U}_{t + \tau}({\boldsymbol\lambda})\ket{0}\} \\
& ~~~~~~~+ l_{2}{\rm Re}\{\bra{0} \hat{U}_{t}^{\dagger}\bigl(\hat{A} - \hat{A}^{\dagger}\bigl){D_{t}^{\dagger}}\hat{U}_{t + \tau}({\boldsymbol\lambda})\ket{0}\} \Bigl]^{2} \;,~~~~
\end{aligned}\end{eqnarray}
where $\ket{\Psi_{t+\tau}({\boldsymbol\lambda})} = \hat{U}_{t+\tau}({\boldsymbol\lambda})\ket{0}$ and $\ket{\Psi_{t}} = \hat{U}_{t}\ket{0}$. The hyperparameter $\Lambda_{t+\tau}$ is eliminated through optimization \cite{Umer2024opt} (also refer to Appendix-\ref{AppSec:Burger}), thereby yielding $\Lambda_{t} = \sqrt{\vert C_{I, t}\vert}$, and $C_{I, t}$ represents the optimized value of the cost function at time instance $t$. The dynamics over an extended period is then analyzed by sequentially optimizing the cost function at each time step. 

\begin{figure*}[tbh]
    \centering
    \sbox6{
        \begin{adjustbox}{width = 0.40\textwidth, height = 0.11\textwidth}
        \begin{quantikz}[row sep={0.8cm,between origins}, wire types={q,n,n,n,n}]
        \lstick{$\ket{0}$} & \gate{H} & & \ctrl{1} & & & \ctrl{1} & \gate{H} & \meter{\langle\sigma_{z}\rangle} \\ 
        &\lstick[4, braces = {black, thick}]{$\ket{0}^{\otimes{n}}$} & \setwiretype{q} & \targ{} & \gate[4]{\hat{U}^{\lambda_{j_{0}}}_{t+\tau}} &  \gate[4]{\hat{U}^{\dagger}_{t}} & \targ{} & \\ 
        & & \setwiretype{q} & & & & &\\ 
        & & \setwiretype{q} & & & & &\\ 
        & & \setwiretype{q} & & & & &
        \end{quantikz}
        \end{adjustbox}
        }
    \sbox7{
        \begin{adjustbox}{width = 0.55\textwidth, height = 0.145\textwidth}
        \begin{quantikz}[row sep={0.8cm,between origins}, column sep={0.9cm, between origins}, wire types={q,n,n,n,n,n,n}]
        \lstick{$\ket{0}$} & \gate{H} & & \ctrl{1} & & \ctrl{1}\gategroup[7, steps = 8, style = {dashed, rounded corners, fill=brown!10, inner xsep = 2pt}, background, label style={label position = above, anchor = north, yshift = -0.2cm, xshift = 2.5cm}]{$\hat{A}$} & & & & & & & & & \ctrl{1} & \gate{H} & \meter{\langle\sigma_{z}\rangle} \\ 
        &\lstick[4, braces = {black, thick}]{$\ket{0}^{\otimes{n}}$} &\setwiretype{q} & \targ{} & \gate[4]{\hat{U}^{\lambda_{j_{0}}}_{t+\tau}} & \targ{} & \ctrl{1} & \ctrl{4} & & & & & \ctrl{4} & \gate[4]{~\hat{U}^{\dagger}_{t}~} & \targ{} & \\ 
        & &\setwiretype{q} & & & & \targ{} & & \ctrl{4} & & & \ctrl{4} & & & & \\ 
        & &\setwiretype{q} & & & & & & & \targ{} & \ctrl{3} & & & & & \\ 
        & &\setwiretype{q} & & & & & & & & \targ{} & & & & & \\ 
        &&&&& \lstick[4, braces = {black, thick}]{$\ket{0}^{\otimes{n-2}}$} & \setwiretype{q} & \targ{} & \control{} & & & \control{} & \targ{} & \\ 
        &&&&&  & \setwiretype{q} & & \targ{} & \ctrl{-3} & \control{} & \targ{} & &
        \end{quantikz}
        \end{adjustbox}
        }
    \sbox8{
        \begin{adjustbox}{width = 0.80\textwidth, height = 0.20\textwidth}
        \begin{quantikz}[row sep={0.8cm,between origins}, wire types={q,n,n,n,n,n,n,n,n,n,n}]
        \lstick{$\ket{0}$} & \gate{H} & & \ctrl{1} & & & & & & & \ctrl{7} & \ctrl{1} & & & & & & & & & \ctrl{1} & \gate{H} & \meter{\langle\sigma_{z}\rangle} \\ 
        &\lstick[4, braces = {black, thick}]{$\ket{0}^{\otimes{n}}$} &\setwiretype{q} & \targ{} & \gate[4]{\hat{U}^{\lambda_{j_{0}}}_{t+\tau}} & \ctrl{6} & & & & & & \targ{} & \ctrl{1} & \ctrl{4} & & & & & \ctrl{4} & \gate[4]{\hat{U}^{\dagger}_{t}} & \targ{} & \\ 
        & &\setwiretype{q} & & & & \ctrl{6} & & & & & & \targ{} & & \ctrl{4} & & & \ctrl{4} & & & \\ 
        & &\setwiretype{q} & & & & & \ctrl{6} & & & & & & & & \targ{} & \ctrl{3} & & & & \\ 
        & &\setwiretype{q} & & & & & & \ctrl{6} & & & & & & & & \targ{} & & & & \\ 
        &&&& \lstick[2, braces = {black, thick}]{$\ket{0}^{\otimes{n-2}}$} & \setwiretype{q} & & & & & & & & \targ{} & \control{} & & & \control{} & \targ{} & \\ 
        &&&& & \setwiretype{q} & & & & & & & & & \targ{} & \ctrl{-3} & \control{} & \targ{} & & \\
        &&& \lstick[4, braces = {black, thick}]{$\ket{0}^{\otimes{n}}$} & \setwiretype{q} & \targ{} & & & & \gate[4]{\hat{U}^{\dagger}_{t}} & \targ{} &  \\
        &&& & \setwiretype{q} & & \targ{} & & & & & \\
        &&& & \setwiretype{q} & & & \targ{} & & & & \\
        &&& & \setwiretype{q} & & & &\targ{} & & &
        \end{quantikz}
        \end{adjustbox}
        }
    \begin{tabular}{cc}
    (a) & ~(b)\vspace{-0.25cm}\\ 
    \vspace{0.1cm}\usebox6 & ~\usebox7\\
    \multicolumn{2}{c}{(c)}\vspace{-0.25cm}\\
    \multicolumn{2}{c}{\usebox8}
    \end{tabular}
    \caption{Low-depth design of quantum circuits for evaluating various components of the Burgers\textquotesingle{} equation cost function. Quantum circuits in panels (a-c) measure the cost function constituents ${\rm Re}\{\bra{0} \hat{U}_{t}^{\dagger}\hat{U}^{\lambda_{j_0}}_{t + \tau}\ket{0}\}$, ${\rm Re}\{\bra{0} \hat{U}_{t}^{\dagger}\hat{A}\hat{U}^{\lambda_{j_0}}_{t + \tau}\ket{0}\}$, and ${\rm Re}\{\bra{0} \hat{U}_{t}^{\dagger}\hat{A}{D_{t}^{\dagger}}\hat{U}^{\lambda_{j_0}}_{t + \tau}\ket{0}\}$, respectively. To evaluate ${\rm Re}\{\bra{0} \hat{U}_{t}^{\dagger}\hat{A}^{\dagger}\hat{U}^{\lambda_{j_0}}_{t + \tau}\ket{0}\}$, and ${\rm Re}\{\bra{0} \hat{U}_{t}^{\dagger}\hat{A}^{\dagger}{D_{t}^{\dagger}}\hat{U}^{\lambda_{j_0}}_{t + \tau}\ket{0}\}$, $\hat{A}$ in panel (b) and (c) is inverted to implement $\hat{A}^{\dagger}$. $H$ is the Hadamard gate, $\hat{A}$ is the adder circuit, and $\hat{U}^{\lambda_{j_0}}_{t+\tau}$ is the unitary ansatz operator with $\lambda_{j} = \lambda_{j_{0}}$, while all other variational parameters remain unchanged. \vspace{-0.2cm}}
    \label{Fig:QNPU_Burgers}
\end{figure*}
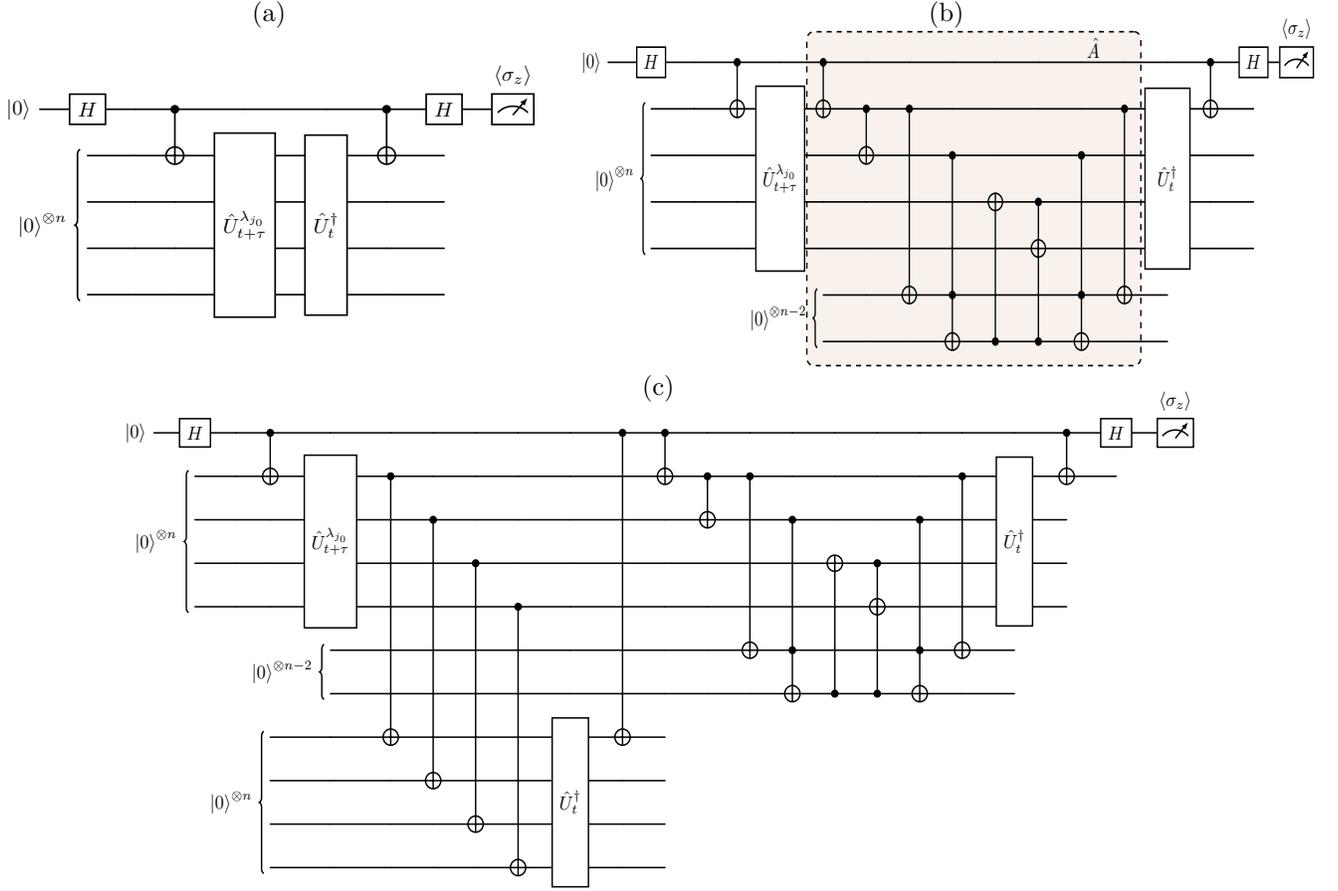

\par To capture the nonlinear dynamics of the fluid velocity field, we adopt the parameterized quantum circuit (PQC) depicted in Fig.~\ref{Fig:Proposed_Ansatz}a, where controlled-$\hat{\tilde{U}}(\lambda_{i})$ operators are defined as controlled-$\hat{R}_{y}(\lambda_{i})$ gates. Each controlled-$\hat{R}_{y}(\lambda_{i})$ gate admits a standard decomposition into two single-qubit rotations and two CNOT gates, as shown in Fig.~\ref{Fig:Proposed_Ansatz}b. To optimize the variational parameters of the PQC, we adopt the sequential grid-based explicit optimization (SGEO) protocol discussed in Ref.~\cite{Umer2024opt}. In this regard, we tailor the cost function to the controlled-$\hat{R}_{y}(\lambda_{i})$ gates of the PQC (refer to Appendix-\ref{AppSec:Burger} for details). The cost function is then expressed as 
\begin{eqnarray}\begin{aligned}
\label{EQ:Burgers3}
&C_{I}({\boldsymbol\lambda}) = \\ &-\Biggl[\biggl(1 + \cos(\lambda_{j}/2) - \sin(\lambda_{j}/2)\biggl)\bigl[G_{1}^{0} + G_{2}^{0} + G_{3}^{0}\bigl] ~~\\ &  + \sin(\lambda_{j}/2)\bigl[G_{1}^{\pi} + G_{2}^{\pi} + G_{3}^{\pi}\bigl] \\ & +
\biggl(1 - \cos(\lambda_{j}/2) - \sin(\lambda_{j}/2)\biggl)\bigl[G_{1}^{2\pi} + G_{2}^{2\pi} + G_{3}^{2\pi}\bigl]\Biggl]^{2} \;, ~~~~ 
\end{aligned}\end{eqnarray}
where $G^{\lambda_{j_{0}}}_{1} = \bigl(\Lambda^{*}_{t} - 2l_{1}^{*}\bigl){\rm Re}\{\bra{0} \hat{U}_{t}^{\dagger}\hat{U}_{t + \tau}^{\lambda_{j_{0}}}\ket{0}\}$, $G^{\lambda_{j_{0}}}_{2} = l_{1}^{*}{\rm Re}\{\bra{0} \hat{U}_{t}^{\dagger}\bigl(\hat{A} + \hat{A}^{\dagger}\bigl)\hat{U}_{t + \tau}^{\lambda_{j_{0}}}\ket{0}\}$, $G^{\lambda_{j_{0}}}_{3} = l_{2}{\rm Re}\{\bra{0} \hat{U}_{t}^{\dagger}\bigl(\hat{A} - \hat{A}^{\dagger}\bigl) \hat{D}^{\dagger}_{t}\hat{U}_{t + \tau}^{\lambda_{j_{0}}}\ket{0}\}$, and $\hat{U}^{\lambda_{j_0}}_{t+\tau}$ is the PQC with $\lambda_{j} = \lambda_{j_{0}}$, while all other variational parameters remain unchanged. Here, terms $G_{i}^{\lambda_{j_{0}}}$\textquotesingle{s} are estimated on quantum hardware and inserted into Eq.~(\ref{EQ:Burgers3}), which is then classically evaluated to yield the values of the cost function along $\lambda_{j} \in [-\pi, \pi)$. We adopt the cost function defined in Eq.~(\ref{EQ:Burgers3}) and sequentially optimize all parameters, following the procedure detailed in Ref.~\cite{Umer2024opt}. 

\par Before discussing the numerical results, we present, in Fig.~\ref{Fig:QNPU_Burgers}, the low-depth implementation of Hadamard test circuits utilized to evaluate the $G_{i}^{\lambda_{j_{0}}}$\textquotesingle{s} terms. First, the $G_{1}^{\lambda_{j_0}}$ term (up to a constant) in Eq.~(\ref{EQ:Burgers3}) represents the overlap of two quantum states, which can be measured using the quantum circuit illustrated in Fig.~\ref{Fig:QNPU_Burgers}a. Here, $\hat{U}_{t}$ represents the unitary operator at time $t$, and $\hat{U}^{\lambda{j_{0}}}_{t+\tau}$ denotes the ansatz at the time instance $t+\tau$, evaluated at $\lambda_{j} = \lambda_{j_{0}}$. The component $G_{2}^{\lambda_{j_0}}$ ($G_{3}^{\lambda_{j_0}}$) comprises two terms that differ only in the operators $\hat{A}$ and $\hat{A}^{\dagger}$. For terms in the $G_{2}^{\lambda_{j_0}}$ ($G_{3}^{\lambda_{j_0}}$), we utilize the quantum circuit depicted in Fig.~\ref{Fig:QNPU_Burgers}b~(Fig.~\ref{Fig:QNPU_Burgers}c). In these circuits, the $\hat{A}^{\dagger}$ operation is implemented by inverting the $\hat{A}$ circuit, which is highlighted with beige color in Fig.~\ref{Fig:QNPU_Burgers}b.

\par We now numerically solve the dynamics of fluid velocity fields in the turbulent regime with kinematic viscosity $\nu = 10^{-3}$ and the initial conditions given as a Gaussian profile with standard deviation $\sigma = 0.3$. The VQA simulations are implemented using Qiskit \cite{Anis2021} platform and $5\times10^{4}$ shots per circuit. It is worth highlighting that although we optimize the cost function for Burgers\textquotesingle{} dynamics, we also evaluate the infidelity $F'(t) = 1 - \vert\braket{\Psi_{\rm classical}(t)|\Psi_{\rm opt}(t)}\vert^{2}$ of the optimized variational state $\ket{\Psi_{\rm opt}(t)}$ with respect to the target classical state $\ket{\Psi_{\rm classical}(t)}$. In this section, we adopt a time step of $\tau = \delta_{x}/10$. At the start of the time evolution, the variational parameters are selected to ensure that the PQC produces a state with high fidelity to the initial velocity field. Particularly, the variational parameters are chosen such that the infidelities at $t = 0$, i.e., the deviation of the initial quantum state from the corresponding classical initial state, are observed to be $7.28\times10^{-6}$, $3.92\times10^{-6}$, and $1.31\times10^{-4}$, by configuring the PQC with $d = 3$, $d = 5$, and $d = 7$ layers for $n = 3$, $n = 4$ and $n = 5$, respectively. Here, $n$ denotes the system size, i.e., the number of qubits used to encode $N = 2^{n}$ grid points. It is important to note that, for a fixed system size ($n$), a fewer number of layers ($d$) can impede the preparation of an initial state with high fidelity to its classical counterpart. Consequently, during state preparation, we also assess the expressivity of the ansatz to ensure that it can accurately encode the initial configuration of the fluid velocity field. 

\begin{figure}[thb]\begin{center}
\includegraphics[clip, trim=0.2cm 0.0cm 4.3cm 0.0cm, width=0.99\linewidth, height=1.15\linewidth, angle=0]{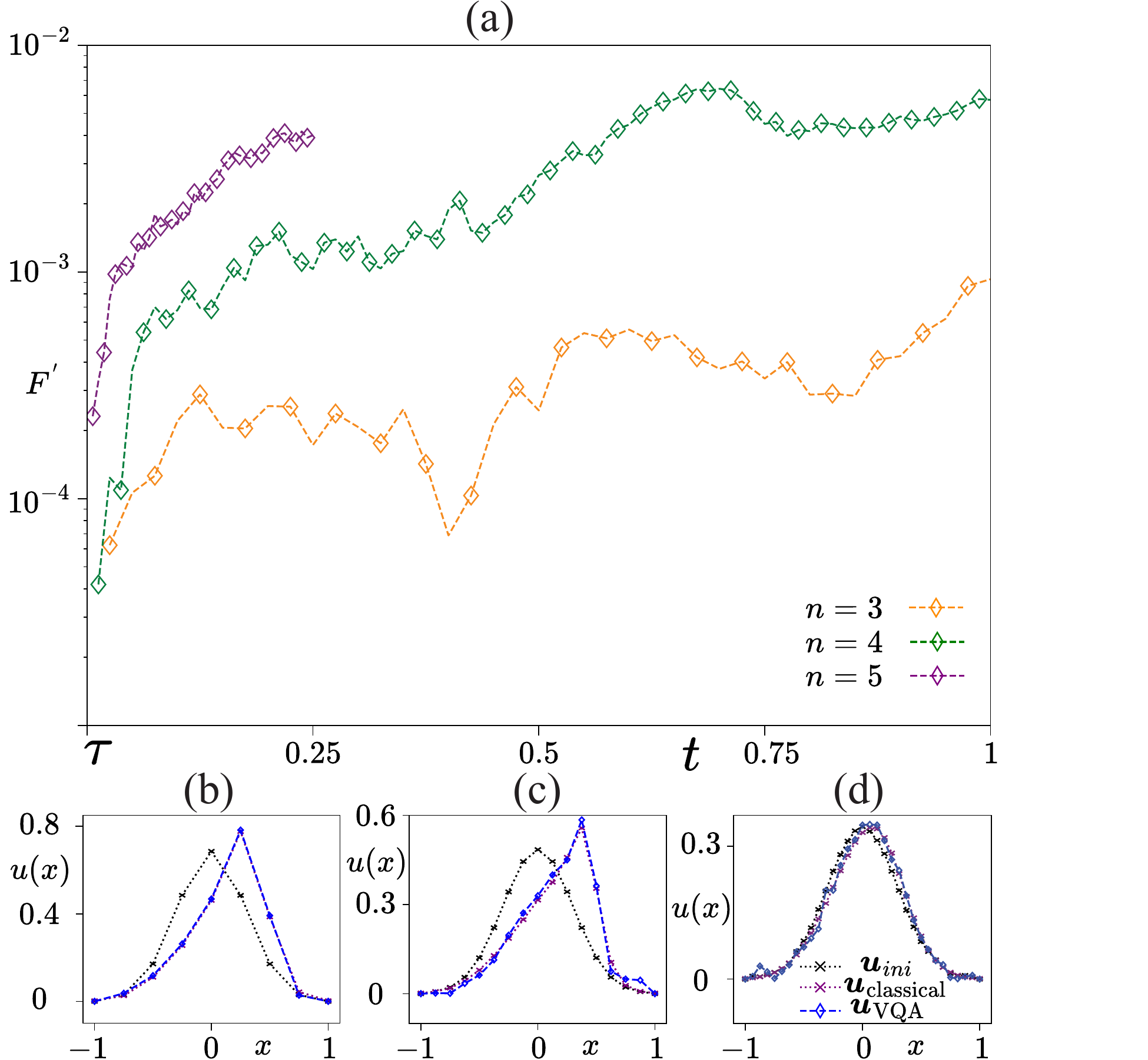}
\caption{Nonlinear dynamics of fluid velocity fields governed by the Burgers\textquotesingle{} equation, with a focus on the analysis of shockwave behavior. Panel (a) illustrates the infidelity $F'(t) = 1 - \vert\braket{\Psi_{\rm classical}(t)|\Psi_{\rm opt}(t)}\vert^{2}$, characterizing evolved states within turbulent regimes with kinematic viscosity $\nu = 10^{-3}$. This metric, $F'(t)$, serves to quantify the discrepancy between states derived from classical simulations and those produced via VQAs using SGEO optimizer. Panel (b), (c), and (d) showcase the fluid velocity field in turbulent regimes for $n = 3$, $n = 4$, and $n = 5$ for $t = 0.7$, $t = 0.9$ and $t = 0.25$, respectively. Here, black color denotes the initial state which is a Gaussian function and purple (blue) color shows the fluid configuration obtained via classical (VQA) simulations. \vspace{-0.2cm}}
\label{Fig:Burgers_Dynamics}
\end{center}\end{figure}

\par With these parameter choices, we perform VQA simulations and present the results in Fig.~\ref{Fig:Burgers_Dynamics}. Specifically, we simulate the dynamics up to $n_{t} = 40$, $n_{t} = 80$, and $n_{t} = 40$ time steps for $n = 3$, $n = 4$ and $n = 5$, respectively. Here, $n_{t}$ denotes the number of time steps. As shown in Fig.~\ref{Fig:Burgers_Dynamics}a, the overlap between the classical and optimized variational states remains above $99\%$ throughout the evolution, corresponding to a state infidelity $F^{'}$ on the order of $10^{-2}$ or lower. Furthermore, the results demonstrate that our low-depth Hadamard test circuits (Fig.~\ref{Fig:QNPU_Burgers}) accurately implement Burgers\textquotesingle{} dynamics over many time steps, and the proposed PQC architecture (Fig.~\ref{Fig:Proposed_Ansatz}a) efficiently captures the accompanying nonlinear fluid behavior. It is worth noting that the overlap exhibits a gradual decrease, typically on the order of $10^{-4}-10^{-5}$ per time step. This behavior may be attributed to the accumulation of statistical noise over time, which gradually degrades the fidelity of the variational state (see Appendix-\ref{AppSec:Shot_Analysis} for detailed analysis). Increasing the number of shots per circuit execution may mitigate this effect by reducing the variance in measurement outcomes, thereby improving the accuracy and stability of the VQA calculations. 

\par To validate our results, we examine the fluid velocity fields at $t = 0.7,~0.9,$ and $0.25$ for systems with $n = 3,~4,$ and $5$ qubits, respectively. In Figs.~\ref{Fig:Burgers_Dynamics}b–\ref{Fig:Burgers_Dynamics}d, the black curves represent the initial state at $t = 0$, while the purple (blue) curves denote the states obtained via classical (VQA) simulations. These results demonstrate an excellent agreement between the classical and variational quantum approaches, as evidenced by the near-perfect overlap of states at distinct time instances. Moreover, key dynamical features such as shockwave formation, manifested as sharp discontinuities in the fluid velocity field, are clearly visible in Figs.~\ref{Fig:Burgers_Dynamics}b and \ref{Fig:Burgers_Dynamics}c. The ability of the VQA framework to capture these nonlinear features substantiates the efficacy of our proposed low-depth Hadamard-test circuits and the PQC architecture.

\subsection{Quantum Gate Resources}
\label{Sec:Gate_Reduction}

\par Having validated the low-depth Hadamard test circuits and assessed the expressivity of our proposed ansatz, we now examine the quantum gate resource requirements for the Burgers\textquotesingle{} dynamics. To this end, we analyze two quantum circuit (QC) architectures: the conventional QCs and their corresponding low-depth QCs. For the conventional QC architecture, we consider the standard construction of the Hadamard test circuits together with the real-amplitude ansatz discussed in our earlier work \cite{Umer2024opt}. Following that study, we set the PQC layers to $d = n$, meaning the number of layers equals the number of qubits, a configuration shown to capture Burgers\textquotesingle{} dynamics with high fidelity (see Ref.~\cite{Umer2024opt} for details). For the low-depth QC architecture, we employ the shallow-depth resource-efficient Hadamard test circuits and the PQC depicted in Fig.~\ref{Fig:QNPU_Burgers} and Figs.~\ref{Fig:Proposed_Ansatz}a-\ref{Fig:Proposed_Ansatz}b, respectively. Guided by, and extrapolated from, the simulation results in Sec.~\ref{Sec:Burgers_Dynamics}, we set the PQC layers to $d = 3, 5, 7, 9, \cdots$ for systems with $n = 3, 4, 5, 6, \cdots$ qubits. In addition to this,  our analysis of gate resource requirements encompasses two hardware platforms: superconducting and trapped-ion. For superconducting hardware, we consider the  IBM\textquotesingle{s} \emph{ibm\_sherbrooke} processor, which has a native gate set $\{{\rm ECR},~R_{z},~X,~sx\}$ and limited qubit connectivity \cite{IBMQ}. For the trapped-ion platform, we consider AQT\textquotesingle{s} \emph{IBEX-Q1} device, which has a native gate set $\{R_{XX},~R_{z},~R\}$ and all-to-all qubit connectivity \cite{AQT_IBEX}. Finally, we employ a fixed transpilation configuration in Qiskit \cite{Anis2021} using the same backend target, circuit optimization level, and layout strategy for all circuits, and therefore obtain deterministic gate counts that allow for a consistent comparison across different implementations.

\begin{figure}[thb]\begin{center}
\includegraphics[clip, trim=0.0cm 3.5cm 0.0cm 0.0cm, width=0.99\linewidth, height=1.0\linewidth, angle=0]{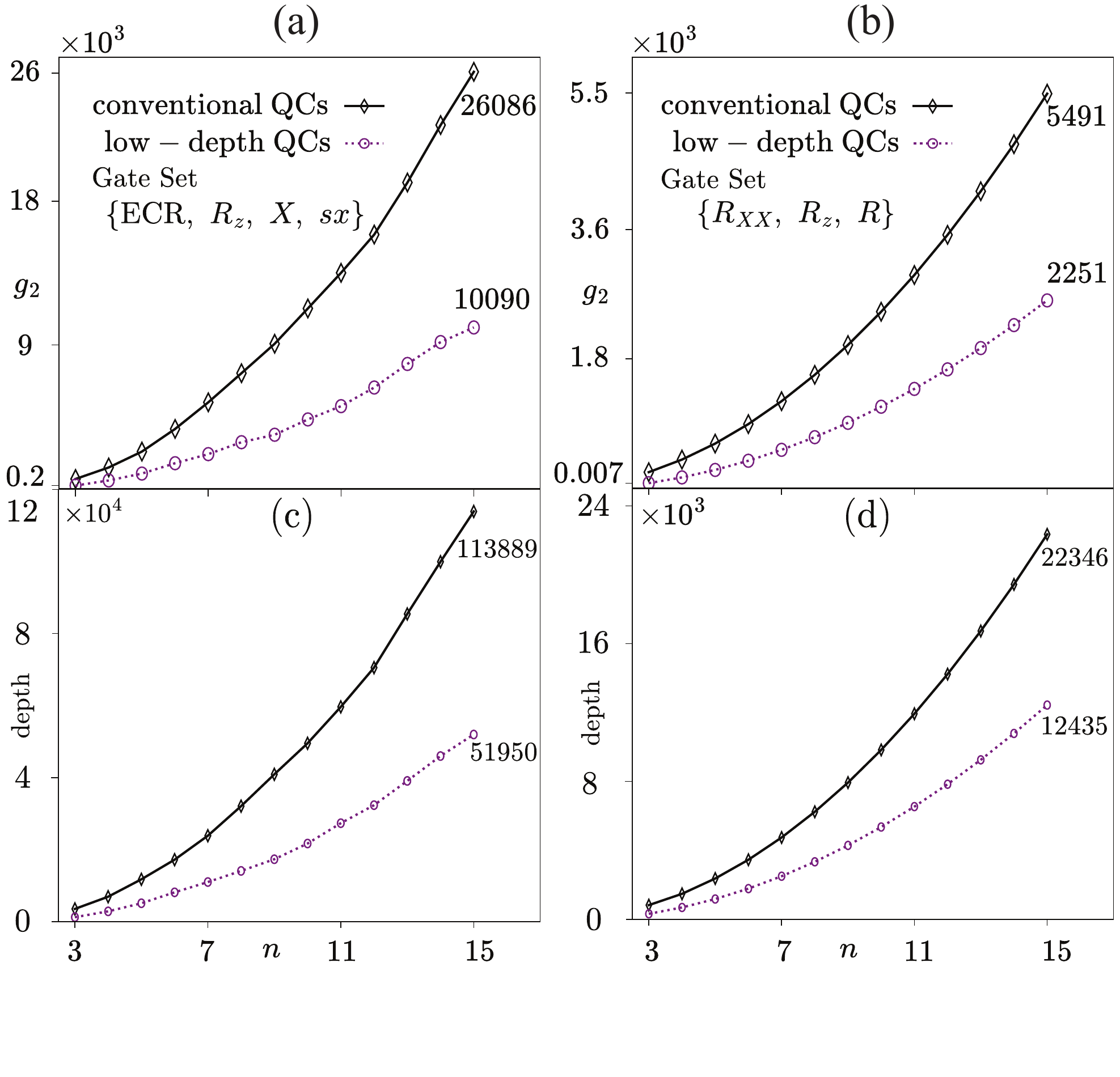}
\caption{Comparison of two-qubit gate count $g_{2}$ and circuit depth for conventional and low-depth Hadamard test QCs for Burgers\textquotesingle{} dynamics. Panel a (b) and c (d) show the two-qubit gate count and circuit depth for superconducting (trapped-ion) based devices, respectively, where black and purple colors indicate the gate count $g_{2}$ for convectional and low-depth Hadamard test QCs. \vspace{-0.2cm}}
\label{Fig:Gate_Count}
\end{center}\end{figure}

\par To elucidate the quantum gate resource requirements of simulating Burgers\textquotesingle{} dynamics, we examine the circuit shown in Fig.~\ref{Fig:QNPU_Burgers}c, which constitutes the most demanding configuration in terms of qubit count, circuit depth, and two-qubit entangling operations. By contrast, the other circuits in Fig.~\ref{Fig:QNPU_Burgers}a and Fig.~\ref{Fig:QNPU_Burgers}b can be executed sequentially with markedly lower qubit and gate requirements. Figure \ref{Fig:Gate_Count} highlights a substantial reduction in the circuit depth and two-qubit gate budget required to simulate Burgers\textquotesingle{} dynamics with our low-depth scheme, compared to the conventional implementation of Hadamard test circuits. Specifically, by analyzing all data points in Fig.~\ref{Fig:Gate_Count}a and Fig.~\ref{Fig:Gate_Count}b, we find that the two-qubit gate count $g_{2}$ is reduced by approximately a factor of two to two and half on both superconducting and trapped-ion hardware platforms. However, despite this reduction in $g_{2}$, the two-qubit gate count in both the unsimplified and simplified circuits continues to scale as $O(n^2)$. At the same time, because of all-to-all qubit connectivity in the trapped-ion processor, they require fewer overall two-qubit gates than their superconducting counterparts, rendering them a particularly favorable platform for executing the variational algorithm. A similar behavior is observed for the circuit depth, as shown in Fig.~\ref{Fig:Gate_Count}c and Fig.~\ref{Fig:Gate_Count}d, where the proposed low-depth construction exhibits consistently smaller depths than the corresponding conventional circuits across the range of problem parameters considered.

\section{Simulations incorporating quantum hardware noise}
\label{Sec:Noisy_Simulations}

\par After reducing the gate counts and circuit depth and assessing the expressive capacity of the parameterized quantum ansatz, we now turn our attention to the performance of noisy simulations of Burgers\textquotesingle{} dynamics. Here, we aim to assess how our efforts enhance the feasibility of implementing the variational algorithm on contemporary NISQ hardware devices. To this end, we retain the previously adopted parameters for Burgers\textquotesingle{} dynamics, except that the kinematic viscosity is fixed at $\nu = 10^{-2}$ in the turbulent regime, the temporal step is set to $\tau = 0.2$, and number of shots are chosen as $2\times{10^{4}}$ per circuit. In addition, we examine a three-qubit system using the ansatz depicted in Fig.~\ref{Fig:Proposed_Ansatz}a, together with the controlled unitary shown in Fig.~\ref{Fig:Proposed_Ansatz}c, employing $d = 3$ layers. Notably, this specific choice of the controlled unitary reduces the number of CNOT gates required per operation compared to a conventional controlled-$R_{y}$ gate. 

\par
\begin{figure}[thb]\begin{center}
\includegraphics[clip, trim=0.1cm 2.1cm 3.0cm 0.0cm, width=0.99\linewidth, height=1.15\linewidth, angle=0]{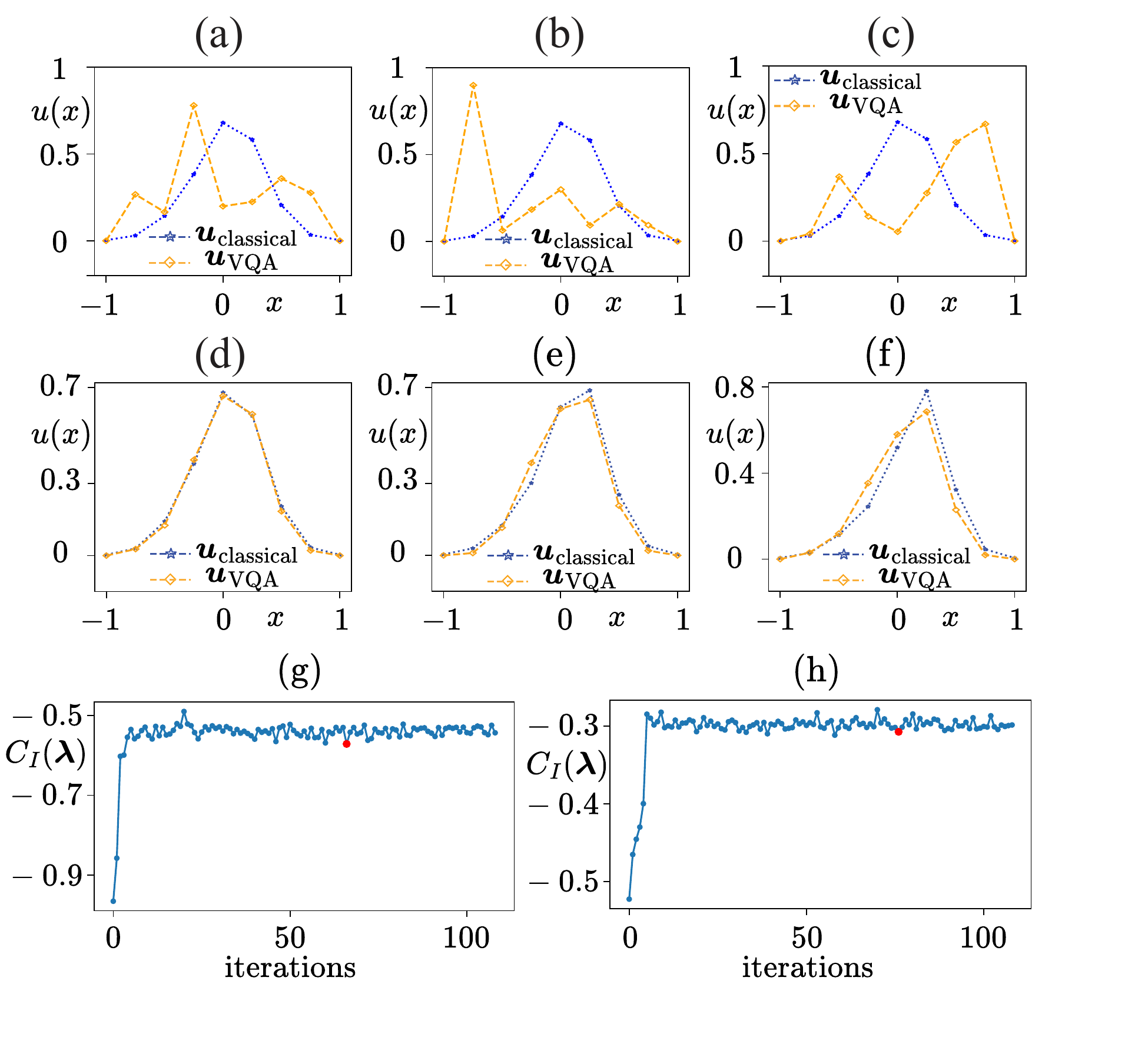}
\caption{Results of simulations in the presence of hardware noise. Panel (a-c) demonstrate the fluid field profile at $t = 0.2$ in the presence of noise models derived from (a) \emph{ibm-brisbane}, (b) \emph{ibm-kingston}, and (c) \emph{ibm-sherbrook} devices. The velocity field ${\boldsymbol{u}}_{\rm VQA}$ has (a) $45.45\%$, (b) $16.63\%$, and (c) $19.26\%$ fidelity with the classical results ${\boldsymbol{u}}_{\rm classical}$. Panels (d-f) show the velocity field at (d) $t = 0.2$, (e) $t = 0.4$, and (f) $t = 0.6$ in the presence of noise model derived from AQT\textquotesingle{s} \emph{IBEX-Q1} device. Here, the fidelity between the variational results ${\boldsymbol{u}}_{\rm VQA}$ and the classical results ${\boldsymbol{u}}_{\rm classical}$ are (d) $99.87\%$, (e) $98.58\%$, and (f) $96.45\%$. Panels (g-h) illustrate the cost function behavior for (g) $t = 0.2$ and (h) t = $0.4$ and the red dots indicate the minimum value of the cost function in the last five optimizer iterations, i.e., five times the total number of variational-parameter updates.\vspace{-0.2cm}}
\label{Fig:Burgers_Nosiy}
\end{center}\end{figure}

 First, we emulate the noise characteristics of the IBM Q superconducting devices within the Qiskit framework. Here, we consider single- and two-qubit gates with (average) error rates on the order of $10^{-4}$ and $10^{-3}$, respectively, and a measurement error probability on the order of
$10^{-2}$. Specifically, we choose the noise properties of the Heron R2 type \emph{ibm-kingston} processor and Eagle R3 type \emph{ibm-brisbane} and \emph{ibm-sherbrook} processors \cite{IBMQ}. Although these devices exhibit similar qubit connectivity (except \emph{ibm-kingston}), single- and two-qubit gate noise are substantially different across these devices (refer to Appendix-\ref{AppSec:IBM_Noise} for detailed noise properties of IBM Q devices). It is worth highlighting that, for this problem size and IBM Q devices, our low-depth scheme has reduced the single- and two-qubit gate counts from $6,374$ and $607$ to $1,868$ and $181$ for \emph{ibm-brisbane} and \emph{ibm-sherbrook} devices and to $806$ and $172$ for \emph{ibm-kingston} device, respectively. This is approximately a three-fold reduction in gate resources. Nonetheless, our simulations indicate that the resulting gate counts remain prohibitively high, despite $>98\%$ two-qubit gate fidelities reported for these IBM Q devices, as evidenced in Fig.~\ref{Fig:Burgers_Nosiy}a-\ref{Fig:Burgers_Nosiy}c by the markedly reduced overlap between the variational state and the classical reference after only a single time step of evolution. The velocity profiles generated as a result of these noisy simulations exhibit random behavior, offering no discernible indication of physical significance. We attribute this behavior to the accumulation of hardware noise arising from restricted qubit connectivity of these IBM Q devices, which necessitates numerous swap operations. The corresponding higher number of noisy two-qubit gates likely degrade the computation and yield the essentially random profiles observed in Fig.~\ref{Fig:Burgers_Nosiy}a-\ref{Fig:Burgers_Nosiy}c. 

\par Second, we emulate the noise properties of AQT\textquotesingle{s} trapped-ion \emph{IBEX-Q1} device \cite{AQT_IBEX}, via Qiskit\textquotesingle{s} \texttt{offline\_simulator\_noise} backend \cite{Anis2021}. Here, the single- and two-qubit gates have average fidelities of $99.97\%$ and $98.7\%$, respectively. Owing to the all-to-all qubit connectivity of trapped-ion device, our low-depth scheme decreases the single- and two-qubit gate counts from $1,244$ and $223$ to $242$ and $43$, respectively. This represents not only more than a fivefold reduction but also a gate count that is substantially lower than that required on IBM Q processors. The corresponding noisy simulations yield encouraging results, as shown in Fig.~\ref{Fig:Burgers_Nosiy}d-\ref{Fig:Burgers_Nosiy}f, where we observe overlaps of $99.87\%$, $98.58\%$, and $96.45\%$ between the variational and classical states at time $t = 0.2$, $t = 0.4$, and $t = 0.6$, respectively. Despite a slight decline in state fidelity, the characteristic turbulent-regime shockwave behavior (discontinuity) starts to develop at $t = 0.4$, which becomes more pronounced at $t = 0.6$. Furthermore, the cost function behavior at $t = 0.2$ in Fig.~\ref{Fig:Burgers_Nosiy}g and $t = 0.4$ in Fig.~\ref{Fig:Burgers_Nosiy}h show that although hardware noise distorts the cost function landscape, the minimum values (red dots) reached during the final five iterations of the optimizer, i.e., five multiplied by the total number of variational-parameter updates, still correspond to variational states with high overlap with the classical benchmarks (Fig.~\ref{Fig:Burgers_Nosiy}d-\ref{Fig:Burgers_Nosiy}e). 

\par Noise models employed in this study approximate quantum hardware error processes by mapping them onto elementary quantum channels, most commonly depolarizing, amplitude damping, and/or phase flip, whose rates are derived from ensemble averaged gate fidelity measurements \cite{Magesan2012}. However, real devices exhibit additional correlated effects, including qubit crosstalk, among others, rendering the actual error landscape considerably more intricate than these overly optimistic noise models. Therefore, after obtaining a preliminary appraisal of algorithmic performance with these simplified noise models, we carry out demonstrations on the physical hardware to establish more faithful performance benchmarks.

\section{Simulation on the Trapped-ion-based quantum device}
\label{Sec:Device_Simulations}

\par Motivated by the encouraging outcomes of the simulations carried out in the presence of noise characteristics of the trapped-ion device, we proceed to a hardware demonstration of our low-depth scheme on the AQT\textquotesingle{s} \emph{IBEX-Q1} processor. The processor consists of $12$ fully connected trapped-ion qubits and implements single- and two-qubit operations with gate fidelities of $(99.97 \pm 0.01)\%$ and $(98.7 \pm 0.3)\%$, respectively. The device features a native gate set comprising the single-qubit rotations $\{R_{Z}, R\}$ and the two-qubit entangling gate $R_{XX}$. Consequently, the resulting two-qubit gate counts for our compiled circuits are expected to align closely with those reported in Fig.~\ref{Fig:Gate_Count}(b). We focus on the same parameter regime outlined in section~\ref{Sec:Noisy_Simulations}, but reduce the number of shots to $500$ per circuit to curb execution time and quantum computational cost. 

\begin{figure}[thb]\begin{center}
\includegraphics[clip, trim=0.0cm 0.2cm 3.4cm 0.0cm, width=0.99\linewidth, height=0.79\linewidth, angle=0]{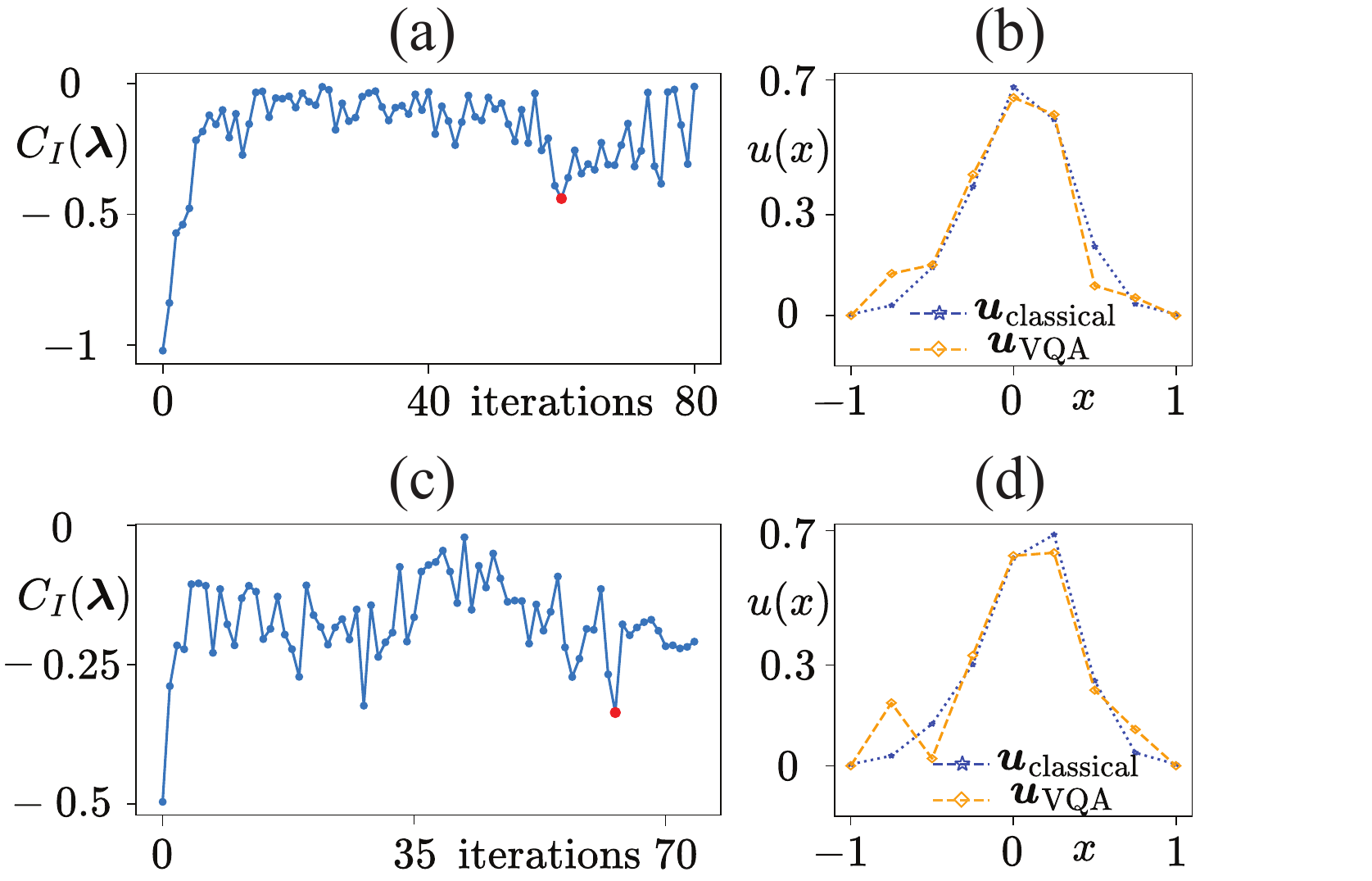}
\caption{Results of the variational algorithm executed on AQT\textquotesingle{s} trapped-ion \emph{IBEX-Q1} processor. Panels (a, c) show the cost function behavior at (a) $t = 0.2$ and (c) $t = 0.4$, where red dot indicate the minimum value for the last five optimizer iterations. Panel (b, d) illustrate the fluid velocity profile at (b) $t = 0.2$ with fidelity $97.48\%$ and (d) $t = 0.4$ with fidelity $95.66\%$. \vspace{-0.2cm}}
\label{Fig:Burgers_IBEX}
\end{center}\end{figure}

\par Figure~\ref{Fig:Burgers_IBEX} illustrates the outcomes of the Burgers\textquotesingle{} dynamics for two consecutive time steps, advancing from $t = 0$ to $t = 0.2$, and subsequently from $t = 0.2$ to $t = 0.4$. It is worth noting that while the variational algorithm was executed on a quantum device to obtain the optimized parameters, the corresponding variational state was subsequently prepared in an ideal (noiseless) setting using these parameters. The variational algorithm converges to a set of parameters that prepare a quantum state exhibiting a $97.48\%$ overlap with the classical velocity field at $t = 0.2$. Moreover, at $t = 0.4$, the quantum state maintains a $95.66\%$ overlap with the corresponding classical profile. The cost function behavior in Fig.~\ref{Fig:Burgers_IBEX}a,~\ref{Fig:Burgers_IBEX}c confirm that, at each time step, the variational minimum coincides with the highest overlap with the classical solution. These observations underscore the importance of our low-depth scheme, which successfully prepares high-fidelity variational states in the turbulent regime of fluid flow on a quantum computer for a small-scale problem. 

\par Low-depth circuit design is paramount for the successful execution of quantum algorithms on NISQ-era hardware, where limited coherence times and imperfect gate fidelities impose strict constraints on circuit depths. As we have shown, shallower circuits with fewer two-qubit gate counts are expected to accumulate fewer stochastic errors, thereby increasing the likelihood that the final state may retain meaningful quantum correlations that faithfully encode the underlying physical phenomena. Collectively, these advantages make low-depth constructions of the quantum circuits the most practical path toward meaningful exploitation of contemporary NISQ processors.

\section{Summary and Outlook}
\label{Sec:Summary}

\par In this work, we proposed a low-depth, resource-efficient construction of a class of Hadamard test quantum circuits. Specifically, we demonstrated that an ancilla-controlled conditional operation, such as a multi-controlled CNOT, can be simplified by removing the ancilla control whenever the gate already includes at least one control on a register qubit. Consequently, the resulting quantum circuits exhibit markedly reduced depth and gate counts. Building on this reduced-depth paradigm, we introduced a parameterized quantum ansatz exclusively designed for the Hadamard test architecture. This ansatz further lowers the gate overhead associated with its parameterized unitary blocks, yielding an even more resource-efficient variational circuit. 

\par To validate the efficacy of our proposed low-depth Hadamard test scheme, we applied it to simulate the nonlinear dynamics of one-dimensional fluid fields governed by Burgers’ equation. In this context, we demonstrated that the proposed parameterized ansatz has sufficient expressivity to capture the Burgers\textquotesingle{} dynamics within the turbulent regime. Specifically, we showed that the ansatz captures the hallmark shockwave behavior of the turbulent regime with a fidelity exceeding $99\%$. Our analysis showed that, relative to conventional Hadamard test circuits, the low-depth scheme achieved an average two-and-a-half-fold reduction in two-qubit gate counts on both superconducting and trapped-ion platforms.

\par To evaluate the effect of hardware noise in low-depth circuit settings, we first examined the variational algorithm in the presence of noise models extracted from calibrated data of contemporary superconducting and trapped-ion processors. These simulations verified that, even in the presence of realistic errors of the trapped-ion hardware, the low-depth scheme preserved overlaps exceeding $96\%$ between the variational and classical target states while accurately reproducing the characteristic shockwave signature. Building on these findings, we demonstrated the execution of the variational algorithm on the AQT\textquotesingle{s} trapped-ion-based \emph{IBEX Q1} processor and observed consistent performance of low-depth circuits, with hardware-generated variational states exhibiting high ($>95\%$) overlap with the corresponding classical targets. Overall, these results verify that our low-depth scheme translated into tangible performance gains on present-day hardware. 

\par In the future, it will be interesting to devise problem-inspired parameterized ans\"{a}tzes tailored to the Hadamard test framework. Such a construction may yield even shallower circuits while simultaneously reducing the parameter count, thereby lowering the computational cost of classical optimization. Additionally, existing low-depth schemes, such as those that leverage ancilla qubits, can be integrated into this approach, thereby further reducing circuit depth. For instance, employing mid-circuit measurement and feedback is expected to replace otherwise deep multi-qubit operations with classically conditioned single-qubit gates, thereby achieving further depth reduction. Moreover, a systematic analysis of various error-mitigation techniques and their impact on the performance of variational algorithms would be particularly worthwhile, especially in the context of our proposed circuit simplifications and ansatz structure. In addition to this, a multi-metric benchmarking of different paradigms, including standard and resource-optimized Hadamard-test schemes, two-copy/SWAP-based estimators, ancilla-free/control-free approaches, and shadow-based or garbage-state methods, remains an important and timely direction for future work. Such a study, which would balance circuit depth and width against measurement and state-preparation overheads as well as classical post-processing costs under matched hardware assumptions, would significantly enhance our understanding of which schemes are most resource-efficient under given hardware conditions. Finally, exploring adaptive, problem-specific ans\"{a}tze tailored to Hadamard test circuits would be valuable, as this strategy is expected to capture the required variational states with even shallower circuits.

\acknowledgments
\par EM and MU proposed the idea of a resource-efficient implementation of quantum circuits, carried out the expressivity analysis of the ansatz, and also performed the analysis under emulated noise models corresponding to IBM Q and AQT devices. MGB, JU, and FR executed the hardware implementation of the problem. FR and DGA supervised the project and guided its overall development. All authors contributed equally to the preparation and writing of the manuscript. 

\par We thank Jos{\'e} Diogo da Costa Jesus for suggesting the use of the controlled-unitary depicted in Fig.~\ref{Fig:Proposed_Ansatz}(c) in place of a controlled-$R_{y}$ gate. We also thank Florian Girtler for valuable discussions during the hardware implementation of the problem. This work was supported by the EU HORIZON—Project 101080085 - QCFD and the National Research Foundation, Singapore through the National Quantum Office, hosted in A*STAR, under its Centre for Quantum Technologies Funding Initiative (S24Q2d0009), Quantum Engineering Programme NRF2021-QEP2-02-P02. FR acknowledges support from HPQC, MILLENION, and QCDC. 

\bibliography{References_QCFD_Burgers_Equation}

\onecolumngrid
\appendix

\section{ Cost function for the Burgers\textquotesingle{} dynamics } \label{AppSec:Burger}

\par We define the square of the residual between the variational state and the time evolved state for the Burgers\textquotesingle{} equation as 
\begin{eqnarray}\begin{aligned}
\label{AppEQ:BurgersCost1}
C_{I} &= ~\vert\vert~\Lambda_{t+\tau}\ket{\Psi_{t+\tau}({\boldsymbol\lambda})} \\ 
&~- \bigl[\Lambda_{t} + l_{1}\big(\hat{A} + \hat{A}^{\dagger} - 2\hat{I}\bigl) - l_{2}\hat{D}_{t}\bigl(\hat{A} - \hat{A}^{\dagger} \bigl)\bigl]\ket{\Psi_{t}}~\vert\vert^{2}\;,~~~~
\end{aligned}\end{eqnarray}
where $l_{1} = {\Lambda_{t}\tau\nu}/{2\delta_{x}^{2}}$, $l_{2} = {\vert\Lambda_{t}\vert^{2}\tau}/{2\delta_{x}}$, $\hat{D}_{t} = {\rm diag}(\psi_{n_{x}, t})$, and $\hat{A}$ is an adder operator that translates the basis states by one unit \cite{Lubasch2020, Jaksch2023}. Here, $\ket{\Psi_{t+\tau}({\boldsymbol\lambda})} = \hat{U}_{t+\tau}({\boldsymbol\lambda})\ket{0}$ and $\ket{\Psi_{t}} = \hat{U}_{t}\ket{0}$ is the state at time $t+\tau$ and $t$, respectively. Simplifying the Eq. (\ref{AppEQ:BurgersCost1}) results in 
\begin{eqnarray}\begin{aligned}
\label{AppEQ:BurgersCost2}
C_{I} &= \vert\Lambda_{t+\tau}\vert^{2} - 2\Lambda_{t+\tau}\Bigl[\bigl(\Lambda_{t}^{*} - 2l_{1}^{*}\bigl)~{\rm Re}\{\bra{0} \hat{U}_{t}^{\dagger}\hat{U}_{t + \tau}({\boldsymbol\lambda})\ket{0}\} \\ 
& ~~+ l_{1}^{*}{\rm Re}\{\bra{0} \hat{U}_{t}^{\dagger}\bigl(\hat{A} + \hat{A}^{\dagger}\bigl)\hat{U}_{t + \tau}({\boldsymbol\lambda})\ket{0}\} \\
& ~~+ l_{2}{\rm Re}\{\bra{0} \hat{U}_{t}^{\dagger}\bigl(\hat{A} - \hat{A}^{\dagger}\bigl){\hat{D}_{t}^{\dagger}}\hat{U}_{t + \tau}({\boldsymbol\lambda})\ket{0}\}\Bigl]~+~{\rm Const.}\;.
\end{aligned}\end{eqnarray}
where Const. is independent of hyperparameter $\Lambda_{t+\tau}$ and variational parameters ${\boldsymbol\lambda}$. Hyperparameter $\Lambda_{t}$ of the instance $t$ is already known during the previous optimization iteration. In fact, we can optimize the cost function with respect to the hyperparameter $\Lambda_{t+\tau}$ by taking $\partial{C_{I}}/\partial{\Lambda_{t+\tau}} = 0$, given $\partial^{2}{C_{I}}/\partial{\Lambda_{t+\tau}^{2}} = 2$ which simplifies the cost function to Eq. (\ref{EQ:Burgers2}) of main text, i.e.,
\begin{eqnarray}\begin{aligned}
\label{AppEQ:BurgersCost3}
C_{I}({\boldsymbol\lambda}) &= - \Bigl[\bigl(\Lambda^{*}_{t} - 2l_{1}^{*}\bigl)~{\rm Re}\{\bra{0} \hat{U}_{t}^{\dagger}\hat{U}_{t + \tau}({\boldsymbol\lambda})\ket{0}\} \\ 
& ~~~~~~~+ l_{1}^{*}{\rm Re}\{\bra{0} \hat{U}_{t}^{\dagger}\bigl(\hat{A} + \hat{A}^{\dagger}\bigl)\hat{U}_{t + \tau}({\boldsymbol\lambda})\ket{0}\} \\
& ~~~~~~~+ l_{2}{\rm Re}\{\bra{0} \hat{U}_{t}^{\dagger}\bigl(\hat{A} - \hat{A}^{\dagger}\bigl){D_{t}^{\dagger}}\hat{U}_{t + \tau}({\boldsymbol\lambda})\ket{0}\} \Bigl]^{2}\;,~~~~
\end{aligned}\end{eqnarray}
where we have ignored the constant term which only shift the magnitude of the cost function toward zero value. 

\par In order to tailor the cost function expression for the SGEO optimizer, we reformulate Eq. (\ref{AppEQ:BurgersCost3}) as follows. First, we express a single controlled-$\hat{W}(\lambda_{i})$ gate as a weighted sum of various terms. Any arbitrary controlled-unitary operation $C\hat{W}(\lambda_{i})$ can be written as  
\begin{eqnarray}\begin{aligned}
\label{EQ:SGEO0}
C\hat{W}(\lambda_{i}) &= \frac{1}{2}\bigl[\hat{I}_{q_{c}}\hat{I}_{q_{t}} + \hat{Z}_{q_{c}}\hat{I}_{q_{t}} + \cos(\lambda_{i}/2)\bigl[\hat{I}_{q_{c}}-\hat{Z}_{q_{c}}\bigl]\hat{I}_{q_{t}} - i\sin(\lambda_{i}/2)\bigl[\hat{I}_{q_{c}}-\hat{Z}_{q_{c}}\bigl]\hat{W}_{q_{t}}\bigl] \\
&= \frac{1}{2}\bigl[ \bigl(1 + \cos(\lambda_{i}/2)\bigl)\hat{I}_{q_{c}}\hat{I}_{q_{t}} + \bigl(1 - \cos(\lambda_{i}/2)\bigl)\hat{Z}_{q_{c}}\hat{I}_{q_{t}} - i\sin(\lambda_{i}/2)\bigl[\hat{I}_{q_{c}}-\hat{Z}_{q_{c}}\bigl]\hat{W}_{q_{t}}\bigl]
\end{aligned}\end{eqnarray}
where $\hat{W} \in \{\hat{X}, \hat{Y}, \hat{Z}\}$ and ${q_{c}}~(_{q_{t}})$ is the control (target) qubit. It is worth noting that $C\hat{W}(0) = \hat{I}_{q_{c}}\hat{I}_{q_{t}}$, $C\hat{W}(2\pi) = \hat{Z}_{q_{c}}\hat{I}_{q_{t}}$, $C\hat{W}(\pi) = \frac{1}{2}\bigl[C\hat{W}(0) + C\hat{W}(2\pi) - i\bigl[\hat{I}_{q_{c}}-\hat{Z}_{q_{c}}\bigl]\hat{W}_{q_{t}}\bigl]$, and $-i\bigl[\hat{I}_{q_{c}}-\hat{Z}_{q_{c}}\bigl]\hat{W}_{q_{t}} = 2C\hat{W}(\pi) - C\hat{W}(2\pi) - C\hat{W}(0)$. Following this, we can write the controlled-$\hat{W}(\lambda_{i})$ operation as, 
\begin{eqnarray}\begin{aligned}
\label{EQ:SGEO1}
C\hat{W}(\lambda_{i}) &= \frac{1}{2}\bigl[ \bigl(1 + \cos(\lambda_{i}/2) - \sin(\lambda_{i}/2) \bigl)C\hat{W}(0) + \bigl(1 - \cos(\lambda_{i}/2) - \sin(\lambda_{i}/2) \bigl)C\hat{W}(2\pi) \\ &~~~~~~~~~~~~~~ + 2\sin(\lambda_{i}/2)C\hat{W}(\pi) \bigl] \;,
\end{aligned}\end{eqnarray}

\par Similarly, one can show that the fundamental building block of Fig.~\ref{Fig:Proposed_Ansatz}c can be written as
\begin{eqnarray}\begin{aligned}
\label{EQ:SGEO2}
C\hat{\tilde{U}}(\lambda_{i}) &= \frac{1}{2}\bigl[\hat{I}_{q_{c}}\hat{I}_{q_{t}} + \hat{Z}_{q_{c}}\hat{I}_{q_{t}} + \cos(\lambda_{i}/2)\bigl[\hat{I}_{q_{c}}-\hat{Z}_{q_{c}}\bigl]\hat{X}_{q_{t}} - \sin(\lambda_{i}/2)\bigl[\hat{I}_{q_{c}}-\hat{Z}_{q_{c}}\bigl]\hat{Z}_{q_{t}}\bigl]
\end{aligned}\end{eqnarray}
and can be written in the form of Eq. (\ref{EQ:SGEO1}).

\par With these expressions, we can formulate the cost function for SGEO optimization by inserting Eq. (\ref{EQ:SGEO1}) or Eq. (\ref{EQ:SGEO2}) in Eq. (\ref{AppEQ:BurgersCost3}), which results in Eq. (\ref{EQ:Burgers3}) of the main text. 

\section{Finite Shot Analysis}
\label{AppSec:Shot_Analysis}
\par In this section, we analyze the dependence of the state infidelity on the number of shots. Specifically, we considered $2\times{10^{2}}$, $5\times{10^{2}}$, $2\times{10^{3}}$, $2\times{10^{4}}$, $5\times{10^{4}}$, $2\times{10^{5}}$, and $5\times{10^{5}}$ shots, while keeping all other parameters fixed at $n = 3$, $d = 3$, $\nu = 10^{3}$, and $\tau = \delta_{x}/10$. The same initial state is used throughout this analysis.

\begin{figure}[thb]\begin{center}
\includegraphics[clip, trim=0.0cm 0.0cm 0.0cm 0.0cm, width=0.9\linewidth, height=0.5\linewidth, angle=0]{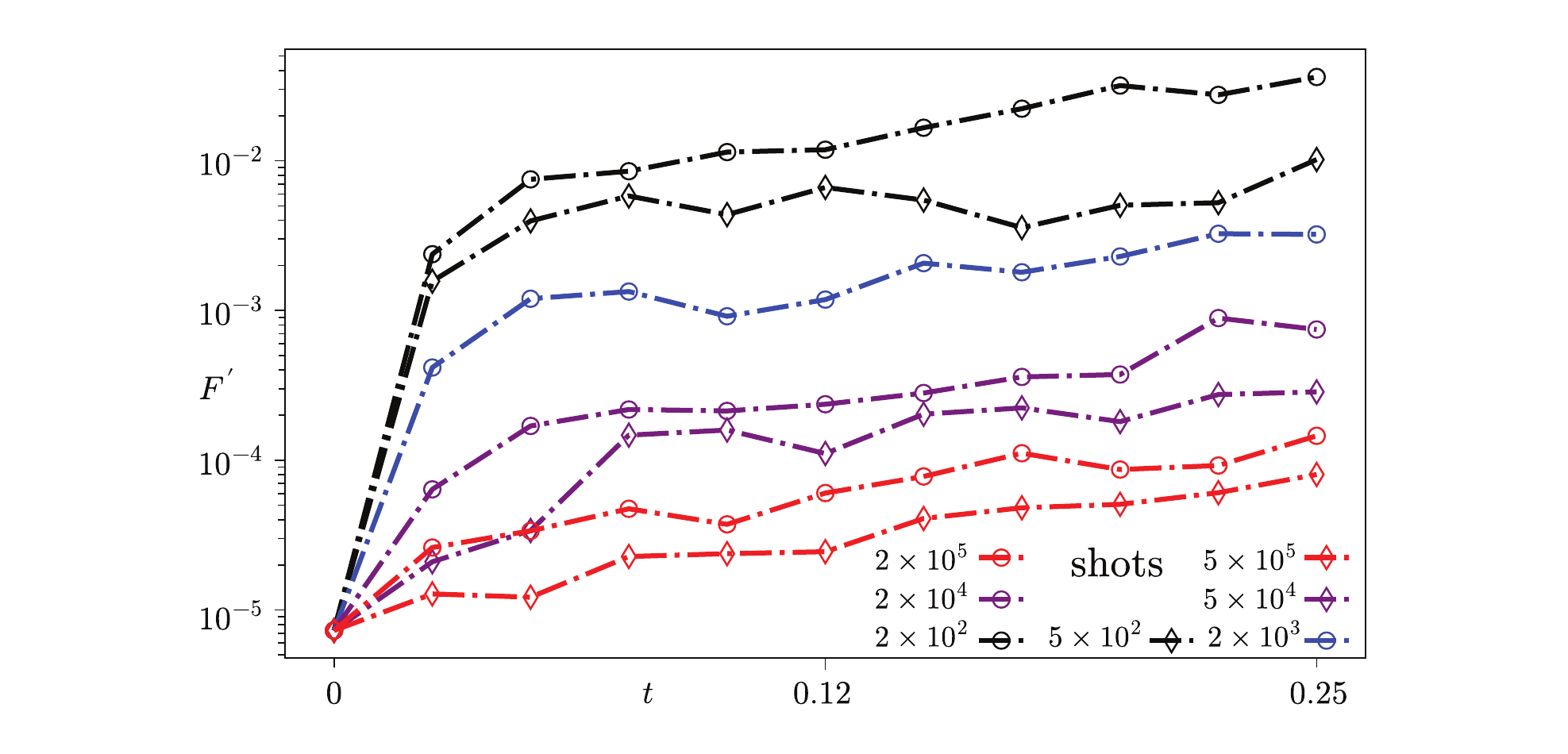}
\caption{Analysis of varying shots on the state infidelities between the variational state and classical target state. Here, we considered the fixed parameters $n = 3$, $d = 3$, $\nu = 10^{-3}$, and $\tau = \delta_{x}/10$. \vspace{-0.2cm}}
\label{Fig:Appendix_Shot_Analysis}
\end{center}\end{figure}

Fig. \ref{Fig:Appendix_Shot_Analysis} highlights that for a fixed initial state, using fewer shots leads to a higher state infidelity during time evolution. After ten time steps, we observe a state infidelity on the order of $10^{-1}$ for $2\times{10^{2}}$ shots, whereas it is reduced to the order of $10^{-4}$ for $5\times{10^{5}}$ shots. This analysis indicates that an increase in the number of shots reduces the infidelity. 

\section{Details on Simulations Incorporating IBM Q Hardware Noise} 
\label{AppSec:IBM_Noise}

\begin{figure}[thb]\begin{center}
\includegraphics[clip, trim=0.0cm 0.0cm 0.0cm 0.0cm, width=0.9\linewidth, height=0.5\linewidth, angle=0]{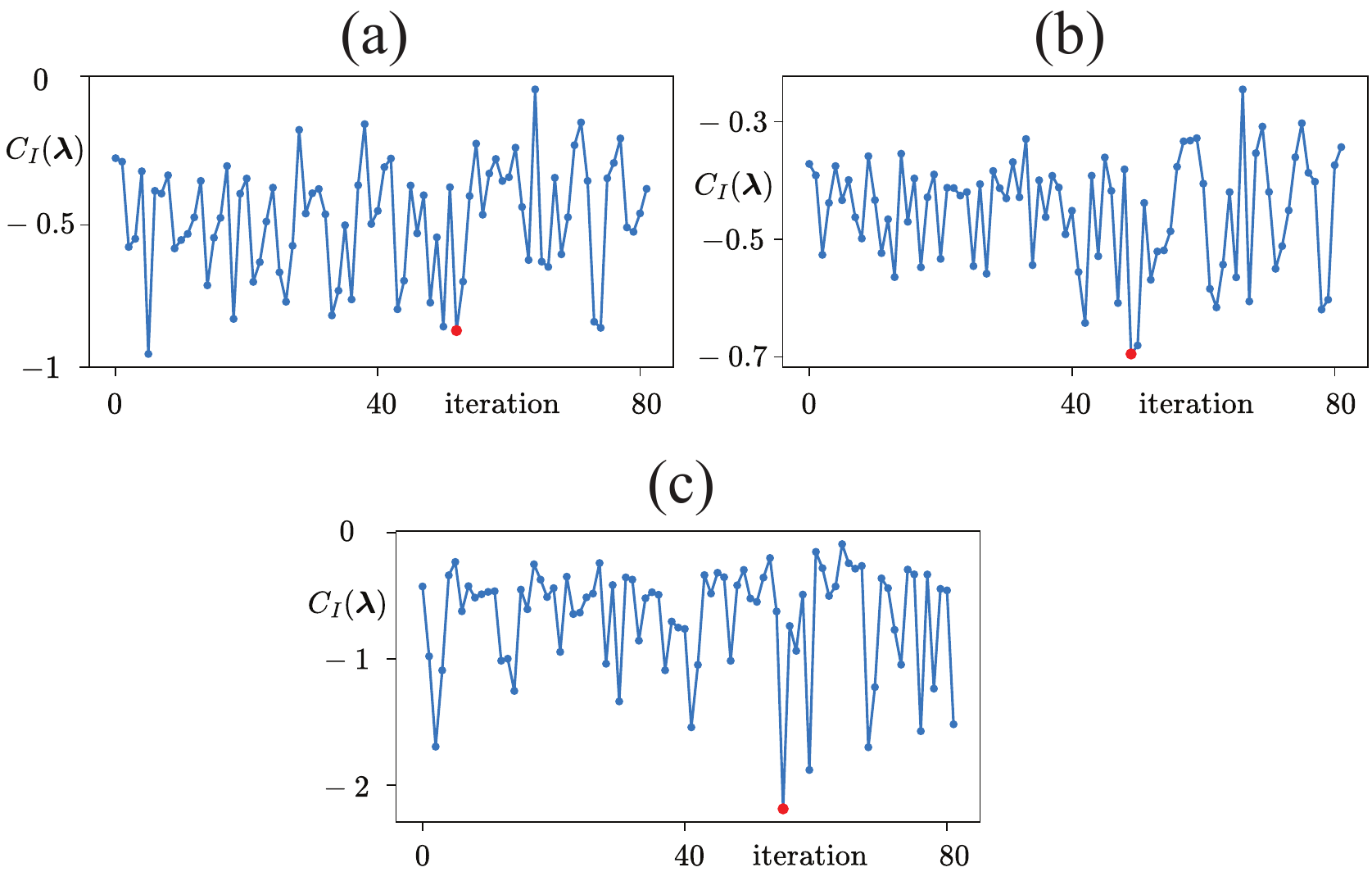}
\caption{Cost function behavior of noisy simulations in the presence of noise model retrieved form (a) \emph{ibm-brisbane}, (b) \emph{ibm-kingston}, and (c) \emph{ibm-sherbrook} devices. Here, red dots indicate the minimum value of the cost function in the last five optimizer iterations. \vspace{-0.2cm}}
\label{Fig:Appendix_Burgers_Nosiy}
\end{center}\end{figure}

In this section, we first discuss the cost function behavior of the noisy simulations incorporating the hardware characteristics of the IBM Q devices. Figures~\ref{Fig:Appendix_Burgers_Nosiy}a-\ref{Fig:Appendix_Burgers_Nosiy}c show the behavior of the cost function vs optimizer iterations, where the red dot indicates the minimum value of the cost function in the last five iterations of the optimizer. We have presented the variational state corresponding to the minimum value of the cost function in Fig.~\ref{Fig:Burgers_Nosiy}a-\ref{Fig:Burgers_Nosiy}c. Figure~\ref{Fig:Appendix_Burgers_Nosiy}a-\ref{Fig:Appendix_Burgers_Nosiy}c reveal that hardware noise severely affect the cost function values, producing random fluctuations rather than convergence to a definite value.

\par In this section, we provide the noise specifications of the IBM cloud-based quantum devices utilized in this work. The devices employed include \emph{ibm-brisbane}, \emph{ibm-kingston}, and \emph{ibm-sherbrook}. Here, \emph{ibm-brisbane} and \emph{ibm-sherbrook} are $127$ qubit device which share a common qubit connectivity topology \cite{IBMQ}, and utilize the gate set $\{{\rm R}_{\rm z}, {\rm X}, {\rm SX}, {\rm ECR}\}$. The \emph{ibm-kingston} device has $156$ qubits with slightly different qubit connectivity \cite{IBMQ} and has gate set $\{{\rm R}_{\rm z}, {\rm R}_{\rm x}, {\rm X}, {\rm SX}, {\rm CZ}, {\rm R}_{\rm zz}\}$. Detailed device specifications, including coherence times and gate and measurement error rates, are presented in Table \ref{Table:IBM_Noise} and Table \ref{Table:IBM_Kingston}.

\par Table \ref{Table:IBM_Noise} and Table \ref{Table:IBM_Kingston} present the noise specifications of the \emph{ibm-brisbane}, \emph{ibm-sherbrook} and \emph{ibm-kingston}, highlighting the qubits utilized in our simulations. Here $P_{01}$, ($P_{10}$) is the probability of measuring state $\ket{0}$ ($\ket{1}$) when the qubit is initially prepared in the state $\ket{1}$ ($\ket{0}$). Moreover, $R$ is the readout assignment error, $\sqrt{X}, X$ is the single-qubit gate error and $ECR$, ${\rm R}_{\rm ZZ}$, and CZ are the two-qubit gate errors for the corresponding gates. In addition to the gate errors listed in the table \ref{Table:IBM_Kingston} for \emph{ibm-kingston} processor, the remaining ${\rm R}_{\rm zz}$-gate and CZ-gate combinations have the following errors: the ${\rm R}_{\rm zz}$-gate between qubit pairs $(1, 0)$, $(2, 1)$, $(3, 2)$, $(4, 3)$, $(5, 4)$, and $(6, 5)$ have errors of $2.09\times{10^{-3}}$, $3.62\times{10^{-3}}$, $4.07\times{10^{-3}}$, $1.88\times{10^{-3}}$, $2.60\times{10^{-3}}$, and $3.45\times{10^{-3}}$, respectively. The CZ-gate between the same pairs of qubits have errors of $2.08\times{10^{-3}}$, $2.42\times{10^{-3}}$, $2.52\times{10^{-3}}$, $2.13\times{10^{-3}}$, $1.48\times{10^{-3}}$, and $1.46\times{10^{-3}}$.

\begin{table*}[h]
\begin{center}
\caption{Hardware specifications and calibration details for the Eagle R3 type \emph{ibm-brisbane} and \emph{ibm-sherbrook} devices are provided. Although these device consists of $127$ qubits, our study utilized only eight qubits, specifically from $q_0$ to $q_7$ \cite{IBMQ}. }
\label{Table:IBM_Noise}
\begin{tabular}{ || c |  c | c | c | c | c | c | c | c | c | c ||}
\hline
~& $T_1$($\mu${s}) & $T_2$($\mu${s}) & Freq. (GHz) & $A$ (GHz)  & $P_{01}$ (\%) & $P_{10}$ (\%) & $R$~(\(10^{-2}\)) & $\sqrt{X}$, X ($10^{-4}$) & Connectivity & $ECR$ ($10^{-3}$) \\ 
\hline
& \multicolumn{9}{c}{\emph{ibm-brisbane}} & \\
 \hline
  0 & 230.13 & 47.19 & 4.72 & -0.311 & 1.07 & 7.03 & 4.05  & 1.98 & (4, 5) & 4.30 \\
 \hline
 1 & 277.15 & 216.71 & 4.81 & -0.309 & 1.46 & 1.95 & 1.70 & 1.24 & (1, 0) & 3.57 \\
 \hline
 2 & 187.00 & 70.44 & 4.61 & -0.309 & 1.27 & 0.58 & 0.92  & 2.06 & (2, 1) & 4.39 \\
 \hline
 3 & 289.31 & 316.62 & 4.87 & -0.309 & 1.75 & 3.71 & 2.73  & 12.90 & (3, 2) & 13.37 \\
 \hline
 4 & 327.73 & 285.37 & 4.81 & -0.310 & 1.22 & 1.56 & 1.39  & 1.90 & (4, 3)& 25.76 \\
 \hline
 5 & 252.21 & 216.01 & 4.73 & -0.311 & 1.56 & 2.00 & 1.78  & 2.02 & (6, 7) & 4.44 \\
 \hline
 6 & 286.37 & 100.14 & 4.87 & -0.309 & 0.78 & 1.46 & 1.12  & 1.37 & (6, 5) & 5.89\\
 \hline
 7 & 375.57 & 319.88 & 4.96 & -0.307 & 0.83 & 0.92 & 0.87  & 2.08 & (7, 8) & 3.21 \\
 \hline
 & \multicolumn{9}{c}{\emph{ibm-sherbrook}} & \\
 \hline
  0 & 512.8 & 304.55 & 4.63 & -0.313 & 0.73 & 0.92 & 0.83  & 4.22 & & \\
 \hline
 1 & 281.82 & 324.96 & 4.73 & -0.312 & 10.54 & 11.67 & 11.11 & 12.8 & (1, 0) & 14.91 \\
 \hline
 2 & 224.95 & 194.79 & 4.81 & -0.311 & 14.74 & 15.82 & 15.28  & 2.31 & (1, 2) & 6.13 \\
 \hline
 3 & 178.94 & 214.92 & 4.74 & -0.311 & 3.36 &  3.17 & 3.27  & 2.04 & (3, 2) & 4.63 \\
 \hline
 4 & 269.13 & 500.95 & 4.78 & -0.310 & 3.76 & 2.05 & 2.90  & 1.44 & (4, 3) & 4.79 \\
 \hline
 5 & 296.69 & 303.84 & 4.85 & -0.310 & 3.22 & 4.39 & 3.80  & 1.96 & (5, 4) & 3.88 \\
 \hline
 6 & 124.46 & 123.99 & 4.90 & -0.309 & 13.28 & 8.88 & 11.08  & 41.9 & (6, 5) & 71.81 \\
 \hline
 7 & 282.8 & 162.34 & 4.75 & -0.311 & 10.54 & 9.66 & 10.11  & 2.88 & (7, 6) & 100.96 \\
 \hline
\end{tabular}
\end{center}
\end{table*}

\begin{table*}[h]
\begin{center}
\caption{Hardware specifications and calibration details for the Heron R2 type \emph{ibm-kingston} processor are provided. Although these device consists of $156$ qubits, our study utilized only eight qubits, specifically from $q_0$ to $q_7$ \cite{IBMQ}. }
\label{Table:IBM_Kingston}
\begin{tabular}{ || c |  c | c | c | c | c | c | c | c | c | c | c ||}
\hline
~& $T_1$($\mu${s}) & $T_2$($\mu${s}) & $P_{01}$ (\%) & $P_{10}$ (\%) & $R$~($10^{-3}$) & $S_{1}$ $10^{-4}$ & $S_{2}$-gate & ${\rm R}_{\rm zz}$ ($10^{-3}$) & CZ ($10^{-3}$) \\ 
\hline
& \multicolumn{8}{c}{\emph{ibm-kingston}} & \\
 \hline
  0 & 381.83 & 410.94 & 2.19 & 4.88 & 35.40  & 2.93 & (0, 1) & 2.09 & 2.08 \\
 \hline
 1 & 318.63 & 502.68 & 0.83 & 0.73 & 7.81 & 2.77 & (1, 2) & 3.62 & 2.42 \\
 \hline
 2 & 303.25 & 116.85 & 0.43 & 0.58 & 5.12  & 1.07 & (2, 3) & 4.07 & 2.52 \\
 \hline
 3 & 363.83 & 469.64 & 0.97 & 0.58 & 7.81  & 3.92 & (3, 4) & 1.88 & 2.13 \\
 \hline
 4 & 210.52 & 85.45 & 1.12 & 1.90 & 15.14  & 1.77 & (4, 5) & 2.60 & 1.48 \\
 \hline
 5 & 406.03 & 248.17 & 0.73 & 0.34 & 5.37  & 1.14 & (5, 6) & 3.45 & 1.46 \\
 \hline
 6 & 227.77 & 117.05 & 0.92 & 0.83 & 8.78  & 3.58 & (6, 7)  & 8.57 & 6.98 \\
 \hline
 7 & 351.20 & 194.05 & 3.32 & 2.34 & 28.32  & 2.70 & (7, 6) & 8.57 & 6.98 \\
 \hline
\end{tabular}
\end{center}
\end{table*}

\end{document}